\documentclass[11pt,a4paper]{article}

 \usepackage[ukrainian,english]{babel}
 \usepackage[OT1,T2A]{fontenc}
 \usepackage[cp1251]{inputenc}
 \usepackage{amsmath}
\usepackage[pdftex]{graphicx}
\usepackage{amssymb}
\usepackage{color}

\newcommand{\sli}{\sum\limits}

\newcommand{\vk}{\vec{k}}
\newcommand{\vl}{\vec{l}}

\newcommand{\cM}{{\cal{M}}}
\newcommand{\cB}{{\cal{B}}}

\newcommand{\cH}{{\cal{H}}}

\newcommand{\rhok}{{\rho_{\vec{k}}}}
\newcommand{\rhomk}{{\rho_{-\vec{k}}}}
\newcommand{\etak}{{\eta_{\vec{k}}}}

\begin{document}

\binoppenalty=10000
\relpenalty=1000

\begin{center}
\textbf{\Large{Representation of the grand partition function of the cell model: \\ the state equation in the mean-field approximation}}
\end{center}

\vspace{0.3cm}

\begin{center}
M.~Kozlovskii, O.~Dobush\footnote{dobush@icmp.lviv.ua}
\end{center}

\begin{center}
Institute for Condensed Matter Physics of the National Academy of Sciences of Ukraine, Lviv, Ukraine
\end{center}

 \vspace{0.5cm}

 The method to calculate the grand partition function of a particle system, in which constituents interact with each other via potential, that include repulsive and attractive components, is proposed. The cell model, which was introduced to describe critical phenomena and phase transitions, is used to provide calculations. The exact procedure of integration over particle coordinates and summation over number of particles is proposed. As a result, an evident expression for the grand partition function of the fluid cell model is obtained in the form of multiple integral over collective variables. As it can be seen directly from the structure of the transition jacobian, the present multiparticle model appeared to be different from the Ising model, which is widely used to describe fluid systems.
 The state equation, which is valid for wide temperature ranges both above and below the critical one, is derived in mean-field approximation. The pressure calculated for the cell model at temperatures above the critical one is found to be continuously increasing function of temperature and density. The isotherms of pressure as a function of density have horizontal parts at temperatures below the critical one.

\vspace{0.5cm}


\vspace{0.5cm}

\renewcommand{\theequation}{\arabic{section}.\arabic{equation}}
\section{Introduction}
\setcounter{equation}{0}

The problem of microscopic description of a fluid behavior in the vicinity of and below the critical temperature $T_c$ as well as the phase transition of the first order has been attracting attention of scientists
 for over a century and remains not less urgent for today.

Nowadays, most approaches to description of phase transitions and critical phenomena in fluids are based on scaling
ideas, universality hypothesis, renormalization group methods. The following theories are worth mentioning: methods
taking into account the fluctuations within the Van der Waals theory~\cite{Wycz_Anis_04},
field-theoretical approach, which appeared to be very powerful in describing magnetic systems; complete scaling
approach~\cite{PhysRevE.85.031131,Anisimov_CMP_13}, which is essentially phenomenological theory; methods of
integral equations, and in particular self-consistent Ornstein-Zernike approximation
(SCOZA)~\cite{Pini_Stell_Wilding_98,Lee_Stell_Hoye_04}; perturbation series expansion, for example hierarchical
reference theory~\cite{Par_Rea_95,Par_Rea_12}; non-perturbative renormalization group approach~\cite{Caillol_06};
collective variables method~\cite{Yukh_Kol_Idz_13,Yukhn2014}; numerical methods and computer simulations.

The investigation of simple fluids is frequently carried out using the concept of a reference system~\cite{Hansen06}, which is often taken in the form of hard-spheres. The full pair-interaction potential is usually chosen in the form of a function that does not possess the Fourier transform. The hard-spheres potential is itself such a function, as well as widely considered Lennard-Jones potential, or more general Mie potentials. However, in literature there are results for systems of many particles interacting via pair potential possessing the Fourier transform. For instance, the Morse fluid has already been studied: within the integral equation
approach~\cite{apf_11}, by Monte Carlo simulations using both $NpT$ plus test particle method~\cite{okumura_00}
and the grand-canonical transition matrix method~\cite{Singh06}. Usage of such potentials may be sufficient
for some purposes, for example, to describe the liquid-vapor coexistence in liquid metals~\cite{apf_11,Singh06}.
The description of such systems does not need the hard-spheres reference system and, consequently, all the
interaction -- short- and long-ranged -- can be accounted in the framework of a unified approach within the
collective variables method.

This paper is aimed to propose a new method for calculating the grand partition function of a simple fluid using the cell model with interacting potential, which possess the Fourier transform, to state the presence of the phase transition in this model and derive the state equation.

\renewcommand{\theequation}{\arabic{section}.\arabic{equation}}
\section{Problem statement}
\setcounter{equation}{0}

Consider a classical system of $N$ identical particles, that are kept in a volume $V$, and interact via a pairwise additive potential $\Phi \! (r_{ij}),$ where $r_{ij}=|\vec{r}_{i}-\vec{r}_{j}|$ is the distance between i- and j-particle.

The grand partition function (GPF) of this system
%equation (2.1)
\begin{equation}\label{GPF_00}
    \Xi = \sum \limits_{N=0}^{\infty} \frac{z^{N}}{N!} \int \limits_{V} \ldots \int \limits_{V} \exp \left[ - \frac{\beta}{2}\sum \limits_{i,j}^{N} \Phi(r_{ij}) \right] d\vec{r}_{1}\ldots d\vec{r}_{N},
\end{equation}
here $z=\exp(\beta\mu)$ is the activity, $\beta$ is the inverse temperature, $\mu$ is the chemical potential.

In order to calculate Eq. (\ref{GPF_00}) we introduce the cell model, in which each cell is allowed to host an arbitrary number of particles.

The model somehow comparable to one that we propose was considered in~\cite{R1}. The free energy of the latter was considered in~\cite{R2} as an approximation of the correspondent free energy of the continuous system.

So introducing the cell model, we conditionally divide volume $V$ into $N_{v}$ cubes, each of volume
$v=\dfrac{V}{N_{v}}$, and set down the cell interaction potential in the following form
%equation (2.2)
\begin{equation}\label{U_ll}
    \Phi(r_{ij}) = \sum \limits_{\vec{l}_{1}\in \Lambda_{B}} \sum \limits_{\vec{l}_{2}\in \Lambda_{B}} U(l_{12}) \delta (\vec{r}_{i}-\vec{l}_{1})\delta (\vec{r}_{j}-\vec{l}_{2}),
\end{equation}
where $l_{12}=|\vec{l}_{1}-\vec{l}_{2}|$. The Eq. \eqref{U_ll} contains summation over all vectors
 $\vec{l}\in \mathbb{R}^{3}$ in the set $\Lambda_{B}$ with periodic boundary conditions.
%equation (2.3)
 \begin{equation}\label{Llambda}
    \Lambda_{B} = \left\{ \vec{l}=(l^{(1)},l^{(2)},l^{(3)}); l^{(i)}=cm_{i}, m_{i}=0,1,2,\ldots,N_{1}-1 \right\},
 \end{equation}
 here $N_{v}=N_{1}^{3}$ is the total number of cells and $N_{1}$ denotes the number of cells along each axis.

The form of interaction potential defined in Eq. \eqref{U_ll} point that, firstly, cells are able to hold varying number of particles and, secondly, interparticle interaction $\Phi(r_{ij})$ inside a cell is constant, namely it is not subjected by the distance between particles $r_{i}$ and $r_{j}$.  Full lattice interaction potential $U(l_{12})$ is composed of repulsive $\Psi(l_{12})$ and attractive $U_{1}(l_{12})$ parts
 %equation (2.4)
 \begin{equation}\label{gen_pot_r}
U(l_{12})=\Psi(l_{12}) - U_{1}(l_{12}).
\end{equation}

To continue calculations it is convenient to rewrite the expression of potential energy $E=\dfrac{1}{2}\sum \limits_{i,j}^{N} \Phi(r_{ij})$ in the following form
%equation (2.5)
\begin{equation}\label{U_tr_2}
    E = \frac{1}{2}\sum \limits_{\vec{l}_{i},\vec{l}_{j} \in \Lambda_{B}} U(l_{12}) \hat{n}(l_{1})\hat{n}(l_{2}),
\end{equation}
using a functional of the microscopic particle density, which is defined as
%equation (2.6)
\begin{equation}\label{mic_den}
\hat{n}(l)=\sum_{j=1}^N\delta(\vec{r}_{j}-\vec{l}\,),
\end{equation}
here $\vec{r}_j$ is the coordinate of the $j$-particle, $N$ is the number of particles in the system. Imposing periodic boundary conditions the expression for $\hat{n}(\vec{l})$ can be represented
in the form of Fourier series
%equation (2.7)
\begin{equation}\label{n_F}
\hat{n}(l)=\frac{1}{N_{v}}\sum_{\vec{k} \in \cB_{\Lambda}}\hat{\rho}_{\vec{k}}{\rm e}^{{\rm i}\vec{k}\vec{l}},
\end{equation}
here and henceforth the notation $\vec{a}\vec{b}$ should be understood as the dot product of two vectors. The wave vector $\vec{k}$ takes on values from the set $\cB_\Lambda$
%equation (2.8)
\begin{equation}\label{Bk}
    \cB_\Lambda=\left\{ \vk=(k_{x},k_{y},k_{z}) \, \Big| \, k_{i}=-\frac{\pi}{c}+\frac{2\pi}{c}\frac{n_{i}}{N_{1}}, \, n_{i}=1,2,\ldots,N_{1},; \, i=x,y,z; \, N_{v} = N_{1}^{3}   \right\}.
\end{equation}
 The Fourier transform $\hat{\rho}_{\vec{k}}$ of the microscopic particle density
has the form
%equation (2.9)
\begin{equation}\label{ro_hat}
\hat{\rho}_{\vec{k}}=\sum_{j=1}^N\exp(-{\rm i}{\vec{k}}{\vec{l}}_j), \quad {\text{and}} \quad
\hat{\rho}_{\vec{k}=0}=N.
\end{equation}

Now using the Fourier series in Eq. (\ref{n_F}) and the Fourier transform of interaction potential $U_{B}(k)$
%equation (2.10)
\begin{equation}\label{U_tr_3}
   U_{B}(k)= \frac{1}{N_{v}}\sum \limits_{\vec{l}_{12}\in \Lambda_{B}}U_{l_{12}}e^{-i\vec{k}\vec{l}_{12}},
\end{equation}
we can rewrite Eq. (\ref{U_tr_2}) as
%equation (2.11)
\begin{equation}\label{U_tr_4}
     E = \frac{1}{2 N_{v}} \sum \limits_{\vec{k}\in \cB_{\Lambda}} U_{B}(k)\hat{\rho}_{\vec{k}}\hat{\rho}_{-\vec{k}},
\end{equation}
 here $\vec{l}_{12}=\vec{l}_{1}-\vec{l}_{2}$.
Taking into account the latter transformations, the grand partition function (GPF) of the system with the interaction potential $U_{B}(k)$ can be expressed as follows
%equation (2.12)
 \begin{equation}\label{GPF_1}
    \Xi=\sum \limits_{N\geq 0}^{\infty} \frac{z^N}{N!} \int \limits_{V} d\vec{r}_{1} \ldots \int \limits_{V}
d\vec{r}_{N} \exp \left( - \frac{\beta}{2 N_{v}}\sum \limits_{\vec{k}\in \cB_\Lambda } U_B(k)\hat{\rho}_{\vec{k}}\hat{\rho}_{-\vec{k}} \right).
 \end{equation}

The GPF in the collective variables (CV) representation~\cite{Yukh_80e,prepHonop_74e} has the form
%equation (2.13)
\begin{equation}\label{GPF_2}
    \Xi=\sum \limits_{N \geq 0}^{\infty} \frac{z^N}{N!} \int\limits_{V} d\vec{r}_{1} \ldots \int \limits_{V} d\vec{r}_{N} \int (d\rho)^{N_{v}} \exp \left[ - \frac{\beta}{2N_{v}}\sum \limits_{\vec{k}\in \cB_{\Lambda}} U_B(k)\rho_{\vec{k}}\rho_{-\vec{k}} \right] J(\rho - \hat\rho),
 \end{equation}
here
$$
(d\rho)^{N_{v}} = \prod_{\vec{k}\in\cB_{\Lambda}} d\rho_{\vec{k}},
$$
moreover $\rho_{\vec{k}} = \rho_{\vec{k}}^{(c)}-i\rho_{\vec{k}}^{(s)}$, where $\rho_{\vec{k}}^{(c)}$ and $\rho_{\vec{k}}^{(s)}$ are respectively real and imaginary parts.
The function of transition to the CV $\rho_{\vec{k}}$ is a product of delta-functions
%equation (2.14)
 \begin{equation}\label{Jac_gen}
    J(\rho-\hat{\rho})= \prod_{\vec{k}\in \cB_{\Lambda}}\delta(\rho_{\vec{k}}-\hat{\rho}_{\vec{k}}) = \int (d \nu)^{N_{v}} e^{2\pi i \sum\limits_{\vec{k}\in \cB_\Lambda}\nu_{\vec{k}}(\rho_{\vec{k}}-\hat{\rho}_{\vec{k}})},
 \end{equation}
here variables $\nu_{\vk}$ $(\nu_{\vec{k}}=\dfrac{1}{2}(\nu_{\vec{k}}^{(c)}+i\nu_{\vec{k}}^{(s)}))$ are conjugated to $\rhok$, moreover
$$
(d\nu)^{N_{v}} = \prod \limits_{\vec{k}\in\cB_{\Lambda}} d\nu_{\vec{k}}.
$$
The GPF for values of $\vec{k}$, which are discrete and non-restricted from above, was proposed, for example, in~\cite{Yukh_80e} for a many-particle system with Coulomb interaction and was calculated in works~\cite{Yukh_Kol_Idz_13,jukhn_tmf}, using the hard-spheres as a reference system, in the way of adding the hard-sphere repulsive interaction potential to the main interaction potential similar to~(\ref{gen_pot_r}).

In present work, instead of using additional potentials, we select some part from the repulsive component of the interaction potential by means of introducing some value $\chi$
%equation (2.15)
\begin{align}\label{UB_transf}
& \beta U_B(k)= - \beta V(k)+\beta_{c}\chi\Psi(0), \\
& \beta V(k) = \beta U_1(k) - \beta \Psi(k) + \beta_{c}\chi\Psi(0). \nonumber
\end{align}
As a result, we obtain some temperature dependent effective interaction potential $V(k)$ and the reference system potential $\chi\Psi(0)$.  Particles of the reference system interact only if they reside at the same node, and they do it by repulsing each other. Such an interaction is not temperature weighted, therefore it can be used to calculate the jacobian of transition from individual variables to CV.
Thus the effective potential $V(k)$ and the potential of the reference system $\chi\Psi(0)$ now appear in the exponent of Eq. (\ref{GPF_2}) instead of $ U_{B}(k)$. Note that in contrast to~\cite{Yukh_Kol_Idz_13,Yukhn2014} the representation of GPF in Eq. (\ref{GPF_2}) contains a sum over the wave vectors $\vec{k}$, that take on discrete and restricted above values (\ref{Bk}). Let us split the sum over $\vk$ in Eq. (\ref{GPF_2}) into two separate sums for each potential in Eq. (\ref{GPF_2}), replace the value $\rhok$ by $\hat\rho_{\vk}$ in the term with the reference system potential $\chi\Psi(0)$ and use the Stratonovich-Habbard transformation
%equation (2.16)
\begin{equation}\label{Strat-Habb}
    \exp \left[  -\frac{\beta_{c}}{2N_{v}} \sum \limits_{\vk\in\cB_\Lambda} \chi\Psi(0) \hat{\rho}_{\vk}
    \hat{\rho}_{-\vk} \right] =\tilde{g}_{\Psi} \int (d\varphi)^{N_{v}} \exp\left[ -   \frac{N_{v}}{2\beta_{c}} \sum_{\vk\in\cB_\Lambda}\frac{\varphi_{\vk}\varphi_{-\vk}}{\chi\Psi(0)}+ i\sli_{\vk\in\cB_\Lambda}\varphi_{\vk}\hat{\rho}_{\vk}\right],
\end{equation}
%equation (2.17)
 \begin{equation}\label{g_stress}
    \tilde{g}_{\Psi}=\left( 2\pi \frac{\beta_{c}}{N_{v}}\chi\Psi(0)\right)^{-\frac{N_{v}}{2}}.
 \end{equation}
The variables $\varphi_{\vec{k}} = \varphi_{\vec{k}}^{(c)}+i \varphi_{\vec{k}}^{(s)}$ include real $\varphi_{\vk}^{(c)}$ and imaginary
$\varphi_{\vk}^{(s)}$ parts, and also
$$
(d\varphi)^{N_{v}} = \prod \limits_{\vec{k}\in\cB_{\Lambda}} d \varphi_{\vec{k}}.
$$
As a result the GPF Eq. (\ref{GPF_2}) now has the form
%equation (2.18)
\begin{align}\label{GPF_3}
     & \Xi=\tilde{g}_{\Psi}  \sli_{N=0}^\infty \frac{e^{\beta_c\mu^*N}}{N!} \int
     (d\rho)^{N_{v}} e^{\beta \left[\mu-\mu^* (1+\tau)\right]\rho_0} \exp \left[  \frac{\beta}{2N_{v}} \sum \limits_{\vk\in\cB_\Lambda}V(k) \rhok \rhomk \right] \int (d\varphi)^{N_{v}}
     \exp \left[ - \frac{N_{v}}{2\beta_{c}}\sli_{\vk\in\cB_\Lambda}\frac{\varphi_{\vk}\varphi_{-\vk}}{\chi \Psi(0)}\right]  \times
     \nonumber \\
     & \times \int\limits_{V} d\vec{r}_{1} \ldots \int \limits_{V} d\vec{r}_{N}   \exp \left[i \sli_{\vec{k}\in\cB_\Lambda}\varphi_{\vk}\hat{\rho}_{\vk} \right]
     \int (d \nu)^{N_{v}}  \exp \left[2\pi i \sli_{\vk\in\cB_\Lambda} \nu_{\vk}(\rhok-\hat{\rho}_{\vk})\right] .
\end{align}
Here $\mu^{\star}$ is some fixed value of the chemical potential, which is used in the following equality
%equation (2.19)
\begin{equation}\label{exp_mu_star1}
    e^{\beta \mu N}=e^{\beta_{c} \mu^{\star} N} \exp\left[ \beta (\mu-\mu^{\star}(1+\tau)) \hat\rho_{0}\right],
\end{equation}
here $\beta_{c}=(k_{B}T_{c})^{-1}$ is some fixed value of
 inverse temperature, $k_{B}$ is the Boltzmann constant. The identity $\beta_c = \beta(1+\tau)$ is used in Eq. \eqref{exp_mu_star1}, the reduced temperature $\tau$ is expressed here by
 %equation (2.20)
\begin{equation}\label{tau}
\tau=\frac{T-T_{c}}{T_{c}}.
\end{equation}
The Eq. (\ref{GPF_3}) contains the function $J(\rho-\hat\rho)$, which gives a possibility to substitute the quantity $\hat\rho_0$ in Eq. (\ref{exp_mu_star1}) for $\rho_0$ in Eq. (\ref{GPF_3}).

The change of variables
\[
\rhok = \sqrt{N_{v}} \rho'_{\vk}, \quad \nu_{\vk} = \nu'_{\vk} / \sqrt{N_{v}}, \quad \varphi_{\vk} = \varphi'_{\vk} / \sqrt{N_{v}},
\]
as well as introducing the notion  $g_\Psi = \left( 2\pi \beta_{c}\chi\Psi(0)\right)^{-\frac{N_{v}}{2}} $ lead to the following representation of the GPF in the space of CV
%equation (2.21)
\begin{equation}\label{GPF_4}
     \Xi=\int (d\rho)^{N_{v}} e^{\sqrt{N_{v}}\beta \left[ \mu - \mu^* (1+\tau) \right] \rho_0}
     \exp \left[ \frac{\beta}{2} \sum \limits_{\vk\in\cB_\Lambda} V(k) \rho_{\vk} \rho_{-\vk} \right] J(\rho),
\end{equation}
here
%equation (2.22)
\begin{equation}\label{Jac_1}
J(\rho) = \int (d\nu)^{N_{v}} \exp \left[ 2\pi i \sli_{\vk\in\cB_\Lambda}\nu_{\vk}\rhok \right] F(\nu),
\end{equation}
%equation (2.23)
\begin{equation}\label{F_1}
F(\nu) = g_{\Psi} \int (d\varphi)^{N_{v}} \exp \left[ -\frac{1}{2\beta_{c}}\sli_{\vk\in\cB_\Lambda}\frac{\varphi_{\vk}\varphi_{-\vk}}{\chi\Psi(0)}\right] G(\nu).
\end{equation}
The function $G(\nu)$ is the result of summation over the number of particles and integration over their coordinates in the following expression
%equation (2.24)
\begin{equation}\label{G_1}
G(\nu) = \sli_{N=0}^{\infty} \frac{(z^*)^N}{N!}  \int\limits_{V} d\vec{r}_{1} \ldots \int \limits_{V} d\vec{r}_{N} \exp \left[ \frac{- 2 \pi i}{\sqrt{N_{v}}} \sli_{\vec{k}\in\cB_\Lambda} \left(\nu_{\vk} - \frac{\varphi_{\vk}}{2\pi} \right) \hat{\rho}_{\vk}\right].
\end{equation}

\renewcommand{\theequation}{\arabic{section}.\arabic{equation}}
\section{The calculation of the transition jacobian}
\setcounter{equation}{0}

Let us find the evident form of the transition jacobian $J(\rho)$.

 Using the form of the operator $\hat{\rho}_{\vk}$ expressed in Eq. \eqref{ro_hat} we can calculate the Eq. (\ref{G_1}) exactly. As a result, in the site representation
%equation (3.1)
\begin{equation}\label{site}
\nu_l = \frac{1}{\sqrt{N_{v}}} \sli_{\vk\in\cB_\Lambda} \nu_{\vk} e^{-i\vk\vl}, \quad
\varphi_l = \frac{1}{\sqrt{N_{v}}} \sli_{\vk\in\cB_\Lambda} \varphi_{\vk} e^{-i\vk\vl},
\end{equation}
the following expression is obtained
%equation (3.2)
\begin{equation}\label{G_4}
G(\nu) = \exp \left[ \sli_{n=0}^{\infty} \frac{(-2\pi i)^n}{n!} \alpha^{*} \sli_l \Big( \! \nu_l - \frac{\varphi_l}{2\pi} \! \Big)^n \right] = \exp \left[ \alpha^{*} \sli_l e^{-2\pi i \Big( \! \nu_l - \frac{\varphi_l}{2\pi} \! \Big)} \right].
\end{equation}

Evidently, the function $F(\nu)$ in Eq. (\ref{F_1}) takes on the form
%equation (3.3)
\begin{equation}\label{F_2}
F(\nu) \! = \! g_{\Psi}  \int \!\! (d\varphi)^{N_{v}}  \exp \left[-\frac{1}{2\beta_{c}} \sli_{\vk\in\cB_{ \Lambda}}  \frac{\varphi_{ \vk}\varphi_{-\vk}}{\chi\Psi  (0)}\right] \exp \left[\alpha^{ \! *} \!\! \sli_l e^{ \! -2\pi i \Big( \! \nu_l - \frac{\varphi_l}{2\pi} \! \Big)}\right] ,
\end{equation}
here  $\alpha^* = v e^{\beta_c\mu^*} $.

The notion of the Eq. (\ref{F_2}) is symbolic, since $\varphi_{\vk}$ and $\nu_{\vk}$ must be understood as the functions of variables $\varphi_{\vl}$ and $\nu_{\vl}$ according to the Eq. (\ref{site}).

Thus the expression of the transition jacobian can be written in the form
%equation (3.4)
\begin{equation}\label{Ins_1}
    J(\rho)=\prod \limits_{l\in \Lambda} J_{l}(\rho),
\end{equation}
where
%equation (3.5)
\begin{equation}\label{Ins_2}
   J_{l}(\rho)=\int \limits_{-\infty}^{\infty} d \nu_{l} e^{ 2\pi i \nu_{l}\rho_{l} } F_{l}(\nu).
\end{equation}

On the certain stage of calculating $F(\nu)$ (see Appendix A) the equality
\begin{equation}\label{Ins_4}
F_l(\nu) = \sli_{m=0}^\infty \frac{(\alpha^*)^m}{m!} e^{-pm^2} e^{-2\pi i m \nu_l}  \nonumber
\end{equation}
can be obtained, where the parameter $p$ $(p>0)$ can be expressed in the following form
%equation (3.6)
\begin{equation}\label{p}
p = \beta_{c} \chi\Psi(0)/2.
\end{equation}
This parameter is responsible for convergency of the series in the function $F_l(\nu)$ (see Appendix A).
 Taking this into account the jacobian of transition $J(\rho)$ to the collective variables can be written as
%equation (3.7)
\begin{equation}\label{Ins_6}
   J_{l}(\rho)=\sli_{m=0}^\infty \frac{(\alpha^*)^m}{m!} e^{-pm^2} \delta(\rho_{l}-m).
\end{equation}
We can compare the expression Eq. (\ref{Ins_6}) to the transition jacobian of the Ising model
%equation (3.8)
\begin{equation}\label{Ins_7}
    J_I(\rho)= \big[\delta(\rho_{l}+1)+\delta(\rho_{l}-1)\big],
\end{equation}
or the Ising model in the external field $\cH$
%equation (3.9)
\begin{equation}\label{Ins_8}
    J_{I,h}(\rho)= \big[e^{-\beta h}\delta(\rho_{l}+1)+e^{\beta h}\delta(\rho_{l}-1)\big],
\end{equation}
where $h=\mu_{B}\textcolor[rgb]{0.98,0.00,0.00}{\cH}$, $\mu_{B}$ is the Bohr magneton. Easy to see, that the jacobian of the cell model expressed by Eq. (\ref{Ins_6}) is essentially from the transition jacobian of the Ising model. According to the Ising model a cell can either contain a single particle or be empty, as follows from Eq. (\ref{Ins_7}) and Eq. (\ref{Ins_8}). On the other hand each cell of the cell model can host an arbitrary number of particles (this correspond to the multiparticle model).
Moreover the probability to find the m-th particle in the cell is proportional to the term $\exp[-pm^{2}].$

The jacobian of transition to the collective variables $J(\rho)$ can be found from Eq. (\ref{Jac_1}) after calculating the function $F(\nu)$.
The latter can be represented in the form of the following cumulant expansion
%equation (3.10)
\begin{equation}\label{F_cum}
 \bar{F}_{l}(\nu)=  \exp\left[\sum \limits_{n=0}^{m_{0}} \frac{(-2\pi i)^n}{n!} \cM_n \nu_{\vl}^{n}\right].
\end{equation}
The cumulants $\cM_n$ can be expressed using special functions $T_n(\alpha^*, p)$
%equation (3.11)
\begin{equation}\label{T_spec}
T_n(\alpha^*, p) = \sli_{m=0}^\infty \frac{(\alpha^*)^m}{m!}  m^n e^{-pm^2},
\end{equation}
  which have a form of rapidly convergent series, since the parameter $p$ from Eq. (\ref{p}) takes on only positive values, and $\alpha^*$ has been already defined above.

Taking into account the Eq. (\ref{F_cum}), the following expression for the transition jacobian $J(\rho)$ can be obtained
%equation (3.12)
\begin{equation}\label{Jac_3}
    \bar{J}_{l}(\rho_{l})=\exp\left[-\sum \limits_{n=0}^{n_{0}} \frac{a_{n}}{n!}  \rho_{l} ^{n} \right].
\end{equation}
The explicit expressions for coefficients $a_n$, can be found in Appendix A. Note that this coefficients take on only real values.

Thereby the GPF of the cell model has the following form
%equation (3.13)
\begin{equation}\label{GPF_5}
    \Xi= e^{N_{v}\cM_{n}} \int (d\rho)^{N_{v}} exp\Bigg[\sqrt{N_{v}}\beta[\mu-\mu^*(1+\tau)]\rho_0  + \frac{\beta}{2}\sum \limits_{\vec{k}\in \cB_{\Lambda}} V(k) \rhok\rhomk
  - \sum \limits_{n=0}^{\infty}\frac{a_n}{n!} N_{v}^{1-\frac{n}{2}} \sli_{\vec{k}_1,...,\vec{k}_n} \rho_{\vec{k}_1}...\rho_{\vec{k}_n} \delta_{\vec{k}_1+...+\vec{k}_n} \Bigg].
\end{equation}

Apparently, values of the cumulants $\cM_{n}$ and coefficients $a_{n}$ correspond to the use of certain interaction potential,  parameter $p$ (defined by Eq. (\ref{p}) ) and the function $\chi$, that determines the reference system, and also by the temperature $\beta=1/k_{B} T$ (see Appendix A).
To provide further calculations one has to choose particular form of
interaction potential between particles in the system. The expression of GPF
Eq. \eqref{GPF_5} is valid for various potentials composed of attractive and repulsive
parts. Availability of the Fourier transform is the sole restriction to the
class of functions, which are used to model the form of interaction
potential. That is exactly the reason why in present work we do not apply the
reference system made of the hard-sphere potential, which actually was used
to describe fluid systems \cite{Yukhn2014}. We propose the approach of describing fluid system, within which there's no
need to introduce the reference system as a separate object and to use
additional interaction between particles, for example, hard- or soft-sphere
interaction. Exceptionally, to ease mathematical calculations we single out a
term (not dependent on the wave vector) from the Fourier transform of
interaction potential, which is used to count up the jacobian of transition
from individual to collective variables. This term does not change
interaction between particles, that follows from Eq. \eqref{UB_transf}, however it
allows us to obtain the jacobian of transition in compacted form, see Eq.
\eqref{Ins_6}. The latter
is dependent on two parameters: $p$ and $\alpha^*$. The former parameter $p$
is proportional to the constant value $\chi$ and can be determined by the
sort of intermolecular interaction in fluids. The latter parameter $\alpha^*$
defines some fixed value of fugacity $z^* = \exp(\beta\mu^*)$. According to Eq.
\eqref{exp_mu_star1} the value $\mu^*$ can take on arbitrary real values. For this reason,
to provide calculations we choose the value of $\alpha^*=v z^*$, that are
keeping with results represented on Figure~\ref{fig_8}. In general, for each particular
substance we have two parameters: $p$ and $z^*$, that eventually correspond
to critical parameters, for example, critical temperature and critical
density.

To provide further calculations we will choose the interaction potential Eq. (\ref{gen_pot_r}) in the form of the Morse potential, where $U_{1}(l_{12})$ is the attractive part
%equation (3.14)
 \begin{equation}\label{pot_attr}
 U_{1}(l_{12})=2\epsilon e^{-(l_{12}-R_{0})/\alpha},
 \end{equation}
and $\Psi(l_{12})$ is the repulsive component.
%equation (3.15)
  \begin{equation}\label{pot_rep}
  \Psi(l_{12})=\epsilon e^{-2(l_{12}-R_{0})/\alpha}.
\end{equation}
 Here $l_{12}=|\vec{l}_{1}-\vec{l}_{2}|$ is the distance between cell vectors, the value $\epsilon$ determines interaction on the distance $R_0$ between particles,
where the minimal value of $\Phi(l_{12})$ can be reached. The parameter $\alpha$ describes an effective radius of attraction. Widespread use, and multitudinous results from numerical calculations~\cite{apf_11, okumura_00, Singh06} became the main reason stipulated for the choice of Morse potential as the interaction potential $U(l_{12})$ between cells.
The Fourier transform of this potential has the form
%equation (3.16)
 \begin{equation}\label{pot_k12}
U_1(k)=U_1(0)(1+\alpha^{2}k^{2})^{-2}, \qquad \Psi(k)=\Psi(0)(1+\alpha^{2}k^{2}/4)^{-2},
 \end{equation}
here
%equation (3.17)
 \begin{equation}\label{pot_k0}
   U_1(0)=16 \pi \epsilon \left(\frac{\alpha^{3}}{v}\right)e^{R_{0}/\alpha}, \quad    \Psi(0)= \epsilon \pi \left(\frac{\alpha^{3}}{v}\right)e^{2R_{0}/\alpha}.
 \end{equation}

The sign of the value $U(0)=\Psi(0) - U_1(0)$ depends on the parameter
$R_0/\alpha$. For each $\ln 2 < R_0/\alpha < 4\ln 2$ one has $U(0)<0$, and for larger $R_0/\alpha$ the value
$U(0) > 0$.

Summing up the calculations made above, the functional representation of the GPF of a fluid model can be written in frames of "$\rho^{4}$-model" approximation ($n_{0}=4$, see Appendix A) as follows
%equation (3.18)
\begin{align}\label{GPF_5a}
& \Xi= e^{(\cM_0-a_0)N_{v}} \int (d\rho)^{N_{v}} exp\Bigg[\sqrt{N_{v}}\beta[\mu-\mu^*(1+\tau)]\rho_0 - a_1\sqrt{N_{v}}\rho_0 - \frac{1}{2}\sum \limits_{\vec{k}} d(k) \rhok\rhomk - \nonumber \\
  & - \frac{1}{3!} \frac{a_3}{\sqrt{N_{v}}} \sli_{\vec{k}_{1},...,\vec{k}_{3}} \rho_{\vec{k}_{1}}...\rho_{\vec{k}_{3}} \delta_{\vec{k}_{1}+...+\vec{k}_{3}} - \frac{1}{4!} \frac{a_4}{N_{v}} \sli_{\vec{k}_{1},...,\vec{k}_{4}} \rho_{\vec{k}_{1}}...\rho_{\vec{k}_{4}} \delta_{\vec{k}_{1}+...+\vec{k}_{4}}\Bigg],
\end{align}
here $d(k) = a_2 - \beta V(k)$,
and $V(k)$ is the Fourier transform of the effective interaction potential.

We can deal with Eq. \eqref{GPF_5a}, using the method to calculate the GPF of
the Ising model in external field proposed in~\cite{MPK}.
In the present case the value $h = \beta\mu - \beta\mu^* (1+\tau) - a_1$ plays the role of the external field.

\renewcommand{\theequation}{\arabic{section}.\arabic{equation}}
\section{The grand partition function and thermodynamic characteristics}
\setcounter{equation}{0}

The Eq. \eqref{GPF_5a} allows us to calculate the dependence of pressure $P$ on temperature $T$ and chemical potential $\mu$ using the following relation
%equation (4.1)
\begin{equation}\label{state_eq_gen}
   PV=kT \ln{\Xi}.
\end{equation}
The average number of particles $\bar{N}$ can be found if the expression for GPF is known
%equation (4.2)
\begin{equation}\label{aver_particle_num_gen}
    \bar{N}=\frac{\partial \ln{\Xi}}{\partial \beta\mu}.
\end{equation}
The chemical potential can be expressed as a function of either number of particles or relative density, using the Eq. (\ref{aver_particle_num_gen})
%equation (4.3)
\begin{equation}\label{relat_dens1}
 \bar{n}=\frac{\bar{N}}{N_{v}}=\left(\frac{\bar{N}}{V}\right)v,
\end{equation}
where $v$ is a volume of an elementary cell and the parameter of the model in use.

We can use the statements mentioned above and express pressure $P$ as a function of temperature $T$ and relative density $\bar{n}$, by uniting Eq. (\ref{state_eq_gen}) and Eq. (\ref{aver_particle_num_gen}). This function $P = P(\bar{n}, T)$ is actually the state equation of an investigated model.

We will use the latter consideration, but first we have to calculate the GPF.
One of the methods to do this consist in gradual change of variables in Eq. (3.10)
%equation (4.4)
\begin{equation}\label{rok_to_etak}
    \rhok=\etak+n_{c}\sqrt{N_{v}}\delta_{\vk}
\end{equation}
in order to obtain the GPF in the following form
%equation (4.5)
\begin{equation}\label{GPF_6}
\Xi= e^{N_{v}(\cM_0-a_0) + N_{v}E_0(\mu)}\int (d\eta)^{N_{v}} \exp \Bigg[\sqrt{N_{v}} M\eta_0 - \frac{1}{2}\sum \limits_{\vec{k} \in B_{\Lambda}} \tilde d(k) \eta_{\vec{k}}\eta_{-\vec{k}} - \frac{a_4}{4!N_{v}} \sli_{\vec{k}_1,...,\vec{k}_4} \eta_{\vec{k}_1}...\eta_{\vec{k}_4} \delta_{\vec{k}_1+...+\vec{k}_4}\Bigg].
\end{equation}
Here
%equation (4.6)
\begin{align}\label{tilda_coef}
    &M=\beta \mu - \beta \mu^* (1+\tau)-\tilde{a}_{1}, \nonumber \\
    &\tilde{a}_{1} = a_1 + \tilde d(0) n_c + \frac{a_4}{6} n_c^3, \nonumber \\
    &\tilde{d}(k) = \tilde{a}'_{2} - \beta V(k),  \\
    &\tilde{a}'_{2} = a_{2}-n_{c}^{2}\frac{a_{4}}{2}, \nonumber \\
    &n_c = - a_3 / a_4, \nonumber
\end{align}
and the expression of $E_0(\mu)$ is
%equation (4.7)
\begin{equation}\label{E0m}
E_0(\mu)= M n_c + \frac{1}{2} \tilde d(0) n_c^2 + \frac{a_4}{24} n_c^4.
\end{equation}

  An expression for the grand partition function similar to Eq. \eqref{GPF_6} was obtained in~\cite{kozl_2005} for the Ising model in the external
  field. In description of the liquid-gas phase transition the chemical potential acts as the external field. Thus the results obtained
  in~\cite{kozl_2005} can be used here.  Behavior of the order parameter was studied in the set of following works~\cite{kozl_rom_2009,kozl_rom_2010}
  As a result the state equation of the Ising-like model in the external field was obtained. Using the representation of the GPF in the form of
  Eq. (\ref{GPF_6}) the coexistence curve and the spinodal in temperature-magnetization coordinates was obtained
  in~\cite{kozl_rom_2011,kozl_rom_2012}. However, further use of all these results in description of the phase transition in fluid systems need
  additional calculations. Despite the similarity of expressions for the partition function of the Ising model in the external
  field~\cite{kozl_2005,kozl_rom_2009,kozl_rom_2010,kozl_rom_2011,kozl_rom_2012} and the GPF of a fluid Eq. (\ref{GPF_6}),
  there is a main difference in calculating the state equation of systems mentioned above. Thus in case of the Ising model,
  which is the simplest approximation of magnetic system, magnetization is the order parameter. The latter is a function of temperature,
  external field and microscopic parameters of the model~\cite{kozl_rom_2010}. It can be found as a derivative of the free energy with respect
  to external field. The order parameter of the cell model, which is the subject of the present work, is the difference between densities of liquid
  and gaseous phases. On the initial stage of calculations this value is a function of temperature, chemical potential and microscopic parameters
  of the model. Unlike the external field in a magnet, the chemical potential cannot be measured experimentally. The equation
  Eq. (\ref{aver_particle_num_gen}) gives us a possibility to express the chemical potential $M$ from Eq. (\ref{GPF_6}) as a function of the
  density and use this for deducing the state equation Eq. (\ref{state_eq_gen}). As a result of such calculation we obtain the dependence of the
  pressure on temperature and density (instead of temperature and chemical potential, as it could be if formally transfer the results obtained for
  the Ising model in the external field to the fluid model). That's why the calculation of the chemical potential dependence on temperature, density
  and microscopic parameters of the model occupies the central place in deducing the state equation of a fluid.

\renewcommand{\theequation}{\arabic{section}.\arabic{equation}}
\section{The thermodynamic potential of a simple fluid in the mean-field approximation }
\setcounter{equation}{0}

To fulfill the procedure described above we use the following GPF representation of the cell model
%equation (5.1)
\begin{equation}\label{GPF_6a}
\Xi= \sqrt{N_{v}} e^{N_{v}(\cM_0-a_0) + N_{v} E_0(\mu)}Z_{p}\int \limits_{-\infty}^{\infty} (d\rho_{0}) \exp \Bigg[N_{v} E(\rho_{0})\Bigg],
\end{equation}
where
%equation (5.2)
\begin{equation}\label{Zp}
    Z_{p}=\int (d\eta)^{N_{v}-1} \exp \Bigg[- \frac{1}{2}\sum \limits_{\vec{k} \in B_{\Lambda}}' \tilde d(k) \eta_{\vec{k}}\eta_{-\vec{k}} - \frac{a_4}{4!N_{v}} \sli_{\vec{k}_1,...,\vec{k}_4}' \eta_{\vec{k}_1}...\eta_{\vec{k}_4} \delta_{\vec{k}_1+...+\vec{k}_4}\Bigg],
\end{equation}
here the stress sign near sums over the wave vectors means that $\vec{k}\neq 0$. Note that $Z_{p}$ is not a function of the chemical potential,
so the derivative of $Z_{p}$ over the chemical potential is equal to zero. According to this fact contribution of $Z_{p}$ to the pressure
Eq. (\ref{state_eq_gen}) won't depend on the density and the input to the average number of particles $\bar{N}$ will be absent. Taking this into
account we can calculate the GPF in a kind of mean-field approximation (e.g. \cite{kadanoff}), the so called zero mode approximation. Within the
latter we only consider contribution of the wave vector $\vec{k} = 0$ (or more explicitly $\rho_{\vec{k}}=0$ for $\vec{k}\neq 0$; $\rho_0\neq 0$).
Thus the result of calculating Eq. (\ref{GPF_6a}) can be written in the form
%equation (5.3)
\begin{equation}\label{GPF_7}
\ln \Xi_{0} =  N_{v} \left( \cM_0 - a_0 + E_0(\mu)\right) + N_{v} E(\bar\rho_0),
\end{equation}
here $\Xi_0$ denotes the GPF expressed by Eq. (\ref{GPF_6}) in the mean-field approximation
%equation (5.4)
\begin{equation}\label{E0r_1}
E(\bar\rho_0) = M \bar\rho_0 - \frac{1}{2} \tilde d(0) \bar\rho_0^2 - \frac{a_4}{24} \bar\rho_0^4,
\end{equation}
herewith $\bar\rho_0$ is a solution of the equation
%equation (5.5)
\begin{equation}\label{eq_for_ro}
M - \tilde d(0) \bar\rho_0 - \frac{a_4}{6} \bar\rho_0^3 = 0,
\end{equation}
 that leads to the maximal value of $E(\bar\rho_0)$ in Eq. (\ref{E0r_1}).

Calculation of Eq. (\ref{GPF_7}) can be made using the method of steepest descent. According to this method the second derivative
of $E(\bar\rho_0)$ should be
negative, and, consequently, every solution $\bar\rho_0$ should satisfy the condition
%equation (5.6)
\begin{equation}\label{cond_ro00}
\bar\rho_0 > \rho_{00}, \quad \rho_{00} = \left( - \frac{2\tilde d(0)}{a_4}\right)^{1/2}.
\end{equation}
This situation takes place barely in the region $T<T_c$, where $\rho_{00}$ has a real value. In case of $T>T_c$ the equation
Eq. (\ref{eq_for_ro}) has a single solution.

  In order to provide further calculations the critical temperature $T_c$ should be defined. Since we are limited to the mean-field approximation,
  the equation for $T_{c}$ is determined from the following condition
 %equation (5.7)
\begin{equation}\label{D0r_Tc}
\tilde d(0)\Big|_{T=T_{c}}  =0 .
\end{equation}
Using Eq. (\ref{tilda_coef}) we can find corresponding equalities
%equation (5.8)
\begin{equation}\label{kTc}
\beta_c = \frac{\tilde a'_2}{V(0,T_{c})}, \quad k T_c = \frac{V(0,T_{c})}{\tilde a'_2}.
\end{equation}
In case $T\neq T_c$ the following expression appears
%equation (5.9)
\begin{equation}\label{D0r_2}
\tilde d(0) = a'_{2}- \beta V(0).
\end{equation}
The Eq. \eqref{D0r_2} can be transformed to the more ocular form
%equation (5.10)
\begin{equation}\label{D0r_3}
\tilde d(0) = \tilde a_2 \frac{\tau}{1+\tau}, \quad \text{where} \quad  \tilde a_2 = \tilde a'_2 \frac{16e^{-R_{0}/\alpha}-1}{16e^{-R_{0}/\alpha}-1 + \chi}.
\end{equation}

The solution of the equation Eq. (\ref{eq_for_ro}) for $\bar{\rho}_{0}=\bar{\rho}_{0}(\tau, M)$ can be found using the Cardano's method.
The result of this calculation are presented in \emph{Appendix B}.  Thus we can see that single solution for $\bar{\rho}_{0}$ exists in temperature
 region $T > T_{c}$
%equation (5.12)
\begin{equation}\label{1Re_root}
    \bar{\rho}_{0} = \left( \frac{3M}{a_{4}} + \sqrt{Q} \right)^{1/3} +\left( \frac{3M}{a_{4}} - \sqrt{Q} \right)^{1/3},
\end{equation}
here
%equation (5.13)
\begin{equation}\label{discrim}
    Q = \left( \frac{2\tilde{d}(0)}{a_{4}}\right)^{3} + \left( \frac{3M}{a_{4}}\right)^{2}.
\end{equation}
The situation with $\bar{\rho}_{0}$ in temperature region $T < T_{c}$ is a little bit complicated. The number of real solutions
for $\bar{\rho}_{0}$ depends on the value of the chemical potential. For each $|M|>M_{q}$ a single root Eq. (\ref{1Re_root}) exists, here
%equation (5.14)
\begin{equation}\label{MQ}
    M_{q}=\frac{a_{4}}{3}\left(-\frac{2\tilde{d}(0)}{a_{4}}\right)^{\frac{3}{2}}.
\end{equation}
 For $|M|<M_{q}$ the equation Eq. (\ref{eq_for_ro}) has three real solutions
(for more details refer to Appendix B).

\begin{figure}[t!]
\centerline{\includegraphics[width=200pt]{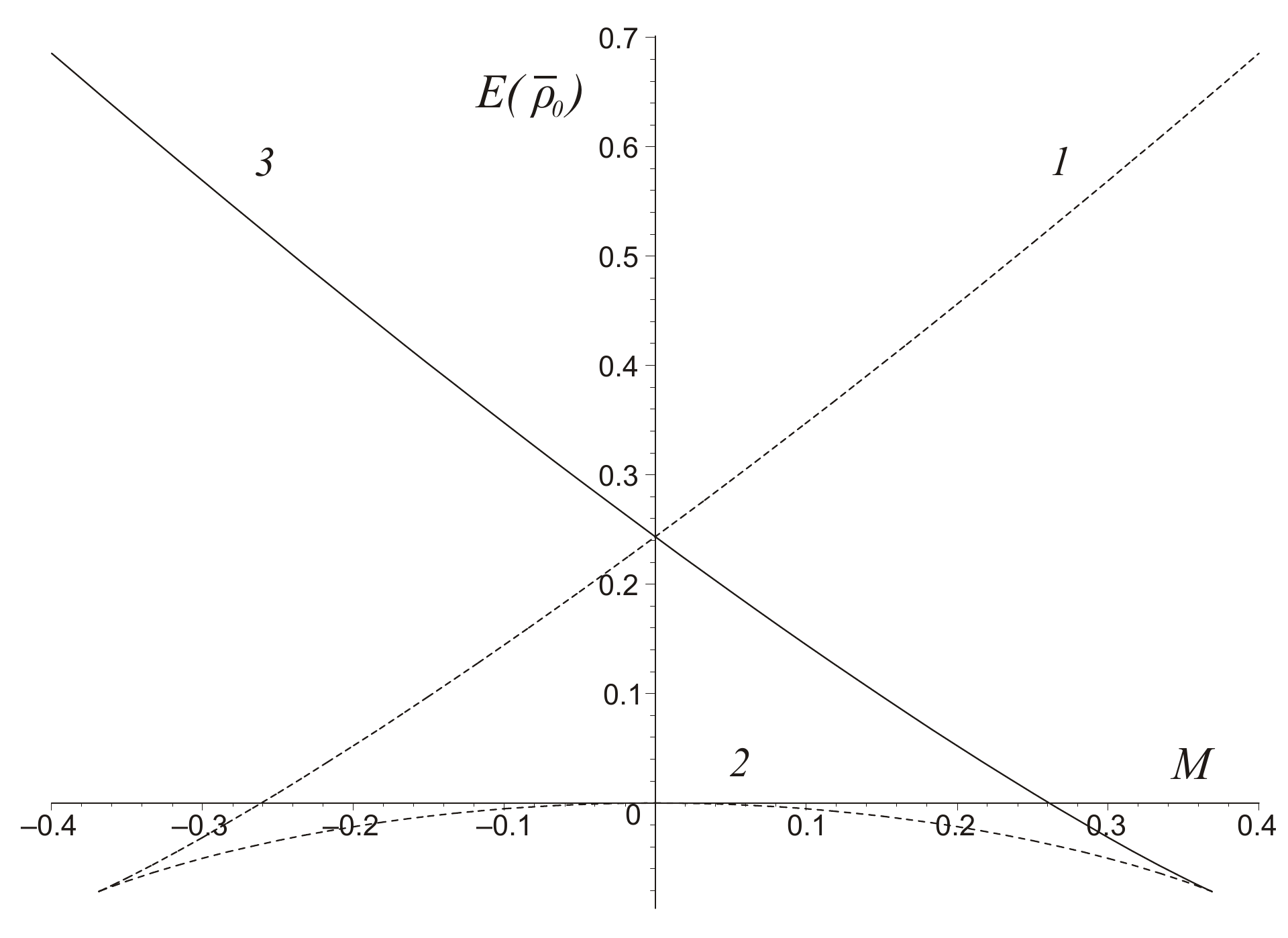}}
\caption{Dependence of the function $E(\bar\rho_{0i})$ on the chemical potential $M$.}\label{fig_1}
\end{figure}

According to Eq. (\ref{state_eq_gen}) the pressure dependence on the temperature and the chemical potential can be found in the following form
%equation (5.15)
\begin{equation}\label{GPF_8}
  \frac{PV}{kT} =  N_{v} \left( \cM_0 - a_0 + E_0(\mu) + E(\bar\rho_{0i}) \right),
\end{equation}
here $E_0(\mu)$ is expressed in Eq. (\ref{E0m}) and the value $ E(\bar{\rho}_{0i})$:
%equation (5.16)
\begin{equation}\label{def_E0r_E0M}
    E(\bar{\rho}_{0i})= M \bar\rho_{0i} - \frac{\tilde{a}_{2}}{2}\frac{\tau}{1+\tau}\bar\rho_{0i}^2 - \frac{a_{4}}{24}\bar\rho_{0i}^4,
\end{equation}
moreover if the Eq. (\ref{eq_for_ro}) has three real solutions, then we should choose the one, which lead to the maximum of the
function $E(\bar{\rho}_{0i})$. The plot of $E(\bar{\rho}_{0i})$ as a function of the chemical potential for each solution $\bar{\rho}_{0i}$ are
represented in  Figure \ref{fig_1}. Note that in order to obtain all the following diagrams we consider the Morse potential with parameters of the
model $R_{0}/\alpha= 3 ln~2$, $\alpha^{*}=15$, $p=0{,}8$ and $\chi=1{,}695$ (for more details refer to Appendix A).

Easy to see, that the solution $\bar{\rho}_{01}$ corresponds to the maximum of $E(\bar{\rho}_{0i})$ for each $M > 0$,
and $\bar{\rho}_{03}$ \-- for $M < 0$.

\begin{figure}[t!]
\centerline{\includegraphics[width=200pt]{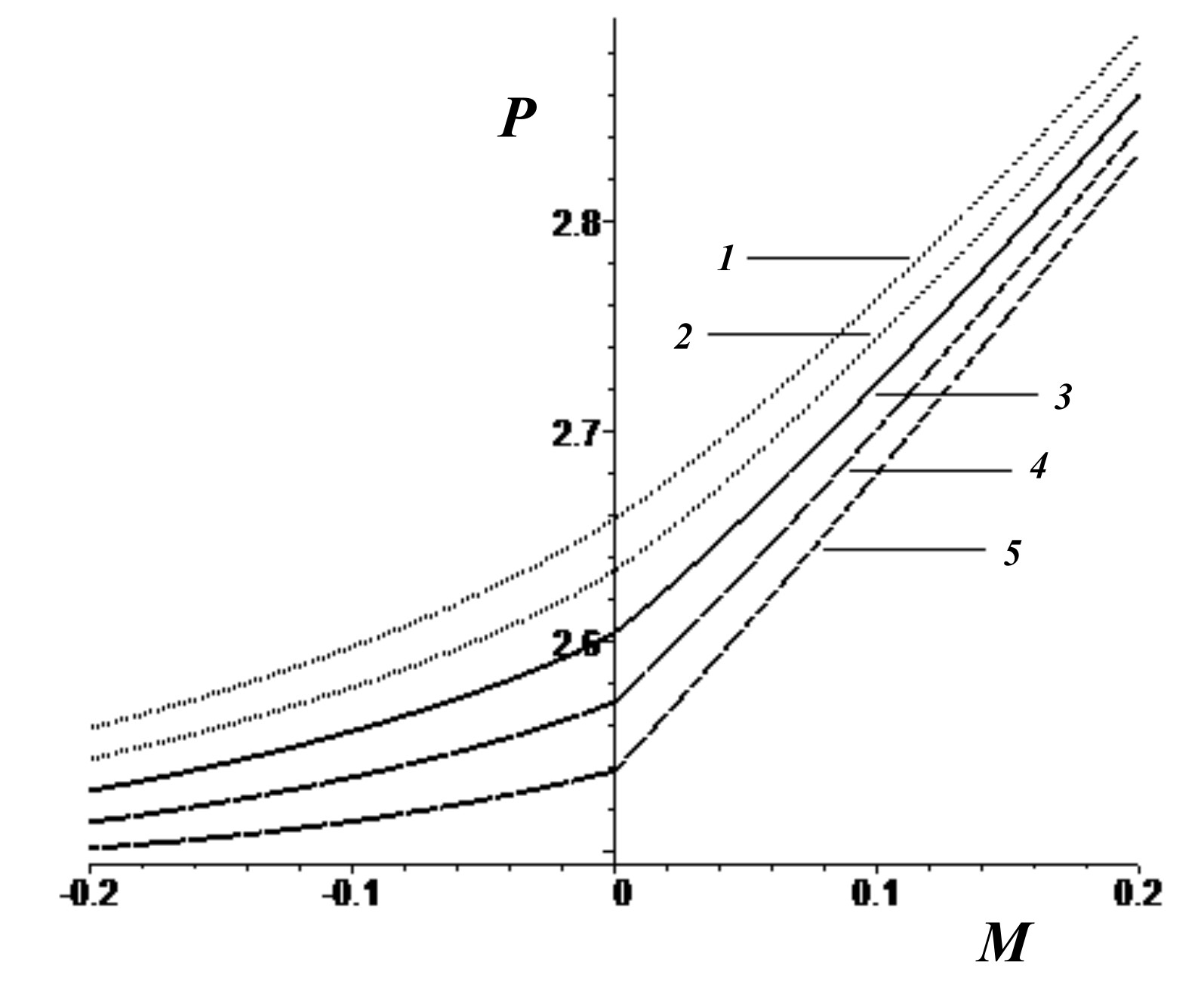}}
\caption{Dependence of the pressure on the chemical potential $M$ at $\tau = 0.2$ (curve 1), $\tau = 0.1$ (2), $\tau = 0$ (3), $\tau = -0.1$ (4), $\tau = -0.2$ (5).} \label{fig_2}
\end{figure}

The curves of pressure as a function of the chemical potential at temperatures above and below the critical temperature are represented in
Figure \ref{fig_2}.

According to the data of estimations the pressure has depend smoothly on the chemical potential at $\tau \geq 0$. For each $\tau < 0$ the
curve $P=P(M)$ has a break at the point $M = 0$. The derivative $\partial P / \partial M$ to the left of the point $M = 0$ is different from the
 same derivative to the right.

\renewcommand{\theequation}{\arabic{section}.\arabic{equation}}
\section{Determining the chemical potential as a function of density and temperature}
\setcounter{equation}{0}

Let us find the average number of particles according to the Eq. (\ref{aver_particle_num_gen}). Taking into account Eq. (\ref{relat_dens1}) the
following expression can be found
%equation (6.1)
\begin{equation}\label{n_nc_ro}
    \bar{n}=n_{c}+\bar{\rho}_{0}.
\end{equation}
Using the value $\bar{\rho}_{0}= \bar{n}-n_{c}$ in the Eq. (\ref{eq_for_ro}), we obtain the equation
%equation (6.2)
\begin{equation}\label{eq_M_n}
    M = \tilde{d}(0)(\bar{n}-n_{c})+\frac{a_{4}}{6}(\bar{n}-n_{c})^{3},
\end{equation}
which link chemical potential with density. This equation enable us to express $M$ via temperature and density and write the state
equation Eq. (\ref{GPF_8}) in the evident form.

Easy to see, that in case $T=T_{c}$ the equality for the chemical potential is as follows
%equation (6.3)
\begin{equation}\label{Mc_1}
    M_{c} = \frac{a_{4}}{6} (\bar{n} - n_{c})^{3}.
\end{equation}
Consequently, the equation of critical isotherm has the form
%equation (6.4)
\begin{equation}\label{state_eq_Tc1}
    \frac{P v}{k T_{c}}=  f_{c} + \frac{a_{4}}{24} \bar n (\bar {n}-n_{c})^3 - \frac{a_{4}}{24} (\bar {n}-n_{c})^4,
\end{equation}
here
%equation (6.5)
\begin{equation}\label{f_c}
    f_{c}= \cM_0 -a_{0} +\frac{a_{4}}{4!}n_{c}^{4}.
\end{equation}
 The expression for critical value of pressure $P_{c}$ can be found at the point $\bar{n}=n_{c}$ as follows
 %equation (6.6)
\begin{equation}\label{Pc}
    P_{c} = \frac{k T_{c} f_{c}}{v}.
\end{equation}

Curves of the function $M=M(\bar{n})$ for different temperatures $T>T_{c}$ is represented on Figure 3.

\begin{figure}[t!]
  \begin{centering}
  \includegraphics[width=200pt]{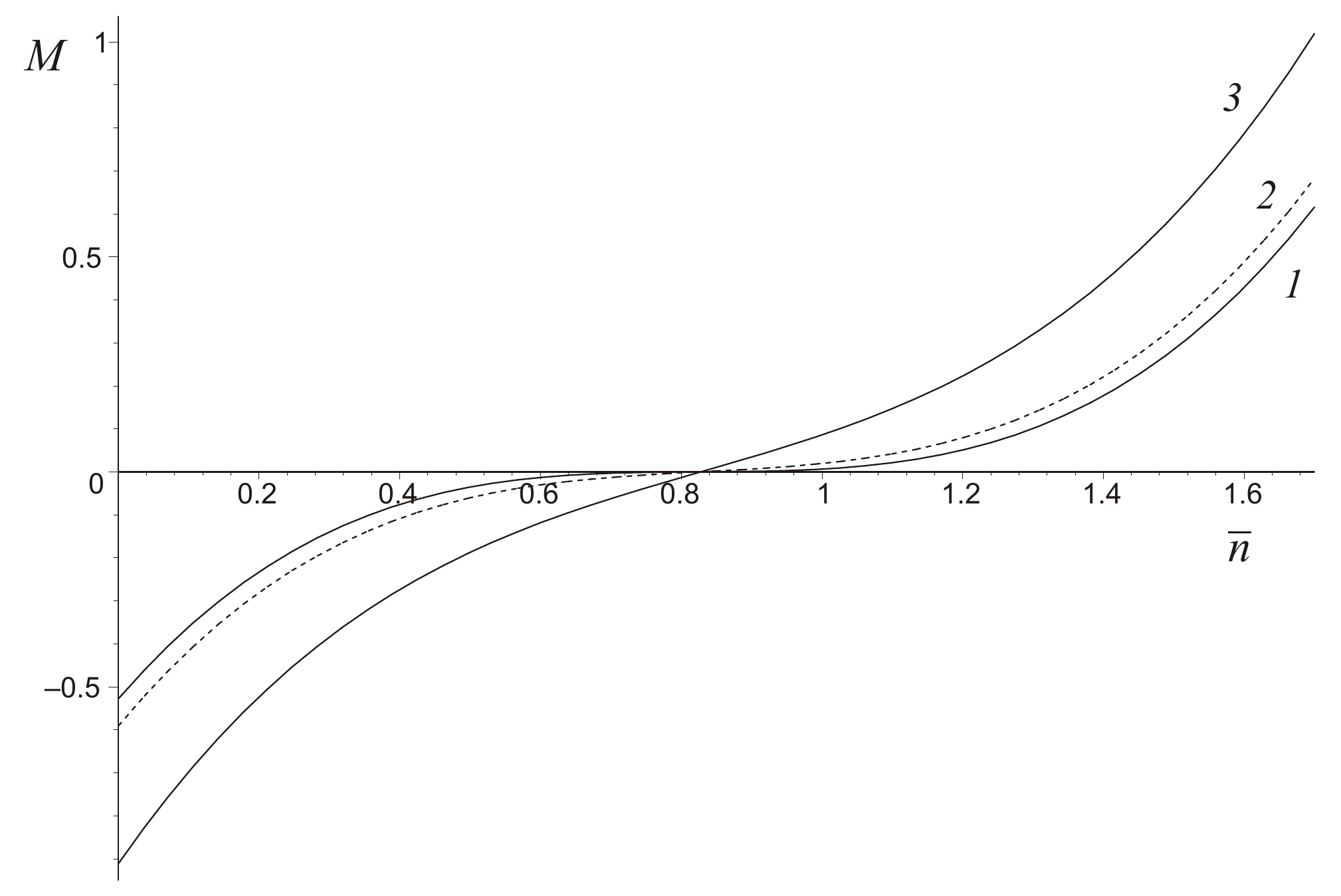}\\
  \caption{Dependence of $M$  on the density $\bar{n}$ at $T>T_c$ for temperatures $\tau_1=0.01$ (curve 1), $\tau_2=0.1$ (curve 2), $\tau_3=1$ (curve 3).}\label{fig_3}
\end{centering}
\end{figure}

The state equation in this temperature region has the form
%equation (6.7)
\begin{equation}\label{state_eq_3}
    \frac{Pv}{kT} = f + \frac{\tilde a_2}{2}  \frac{\tau}{1+\tau}  \bar n^2 + \frac{a_4}{6} \bar n (\bar{n}-n_{c})^{3} - \frac{a_{4}}{24} (\bar{n}-n_{c})^{4},
\end{equation}
here
%equation (6.8)
\begin{equation}\label{fc_Tb}
f = f_c + \frac{1}{2} \tilde d(0) n_c^2.
\end{equation}

\begin{figure}[t!]
 \begin{centering}
  \includegraphics[width=200pt]{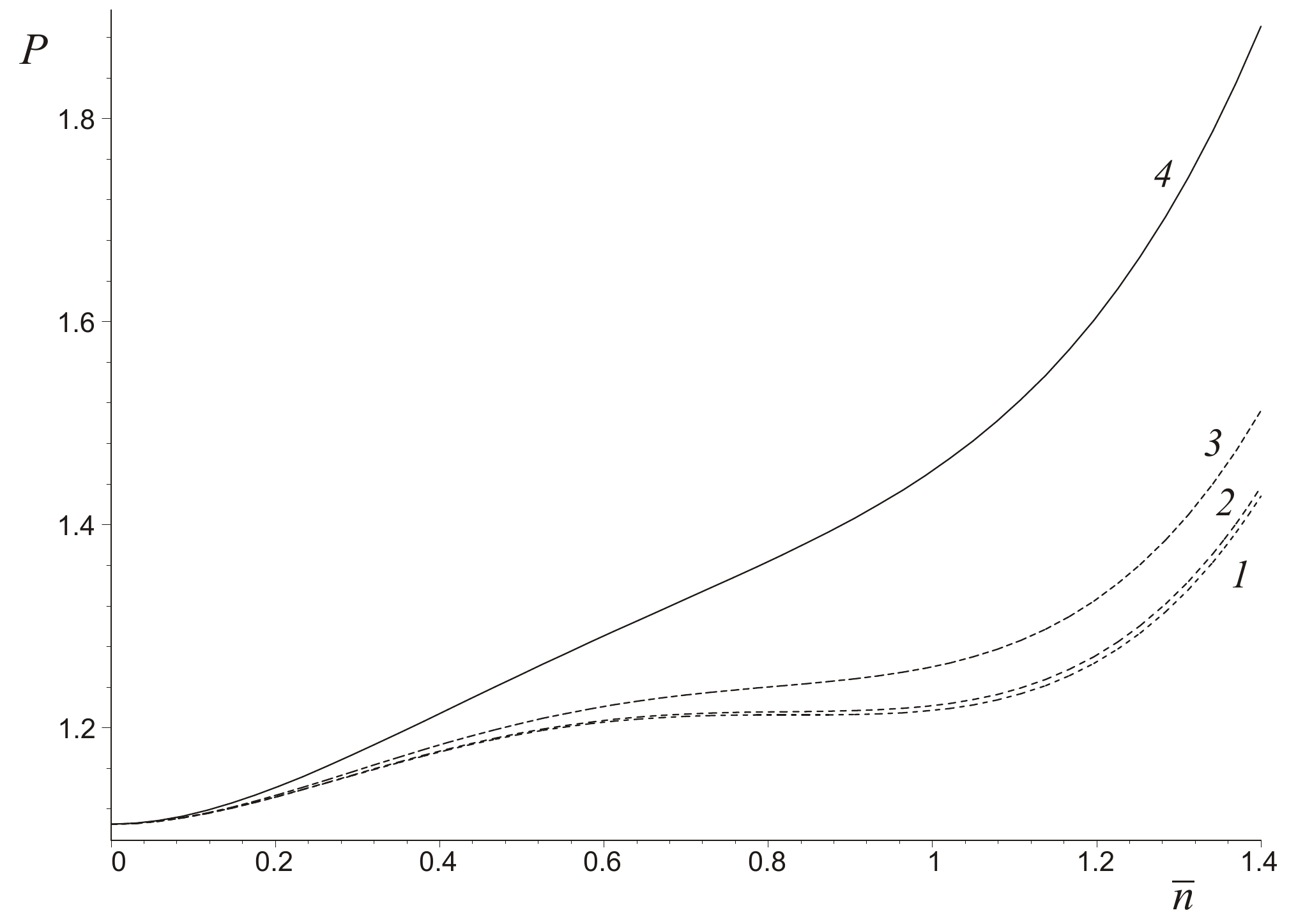}\\
  \caption{Dependence of the pressure $P$ on the density $\bar{n}$ at $\tau=0$  (curve 1), $\tau=0.01$  (curve 2),
  $\tau=0.1$  (curve 3) and $\tau=1.5$  (curve 4).}\label{fig_4}
  \end{centering}
\end{figure}

\begin{figure}[t!]
 \begin{centering}
 \includegraphics[width=200pt]{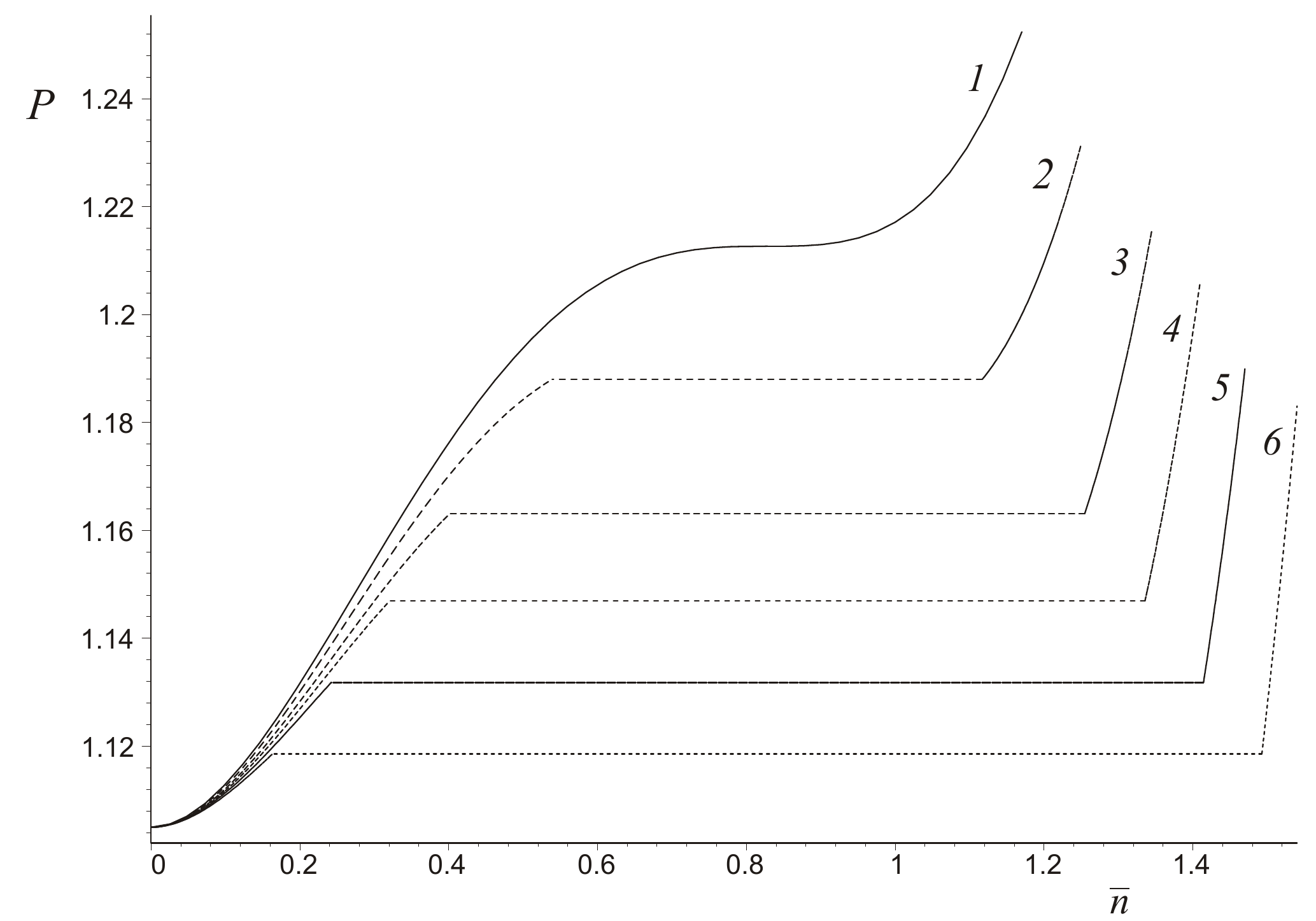}\\
  \caption{Dependence of the pressure $P$ on the density $\bar{n}$ at $\tau=0$  (curve 1), $\tau=-0.07$  (curve 2), $\tau=-0.15$  (curve 3),  $\tau=-0.2$  (curve 4), $\tau=-0.25$  (curve 5), $\tau=-0.3$  (curve 6).}\label{fig_5}
  \end{centering}
\end{figure}

Figures~\ref{fig_4} and \ref{fig_5} depict the pressure dependence on density at $T\geq T_{c}$ and $T\leq T_{c}$ respectively.

\begin{figure}
 \begin{centering}
 \includegraphics[width=200pt]{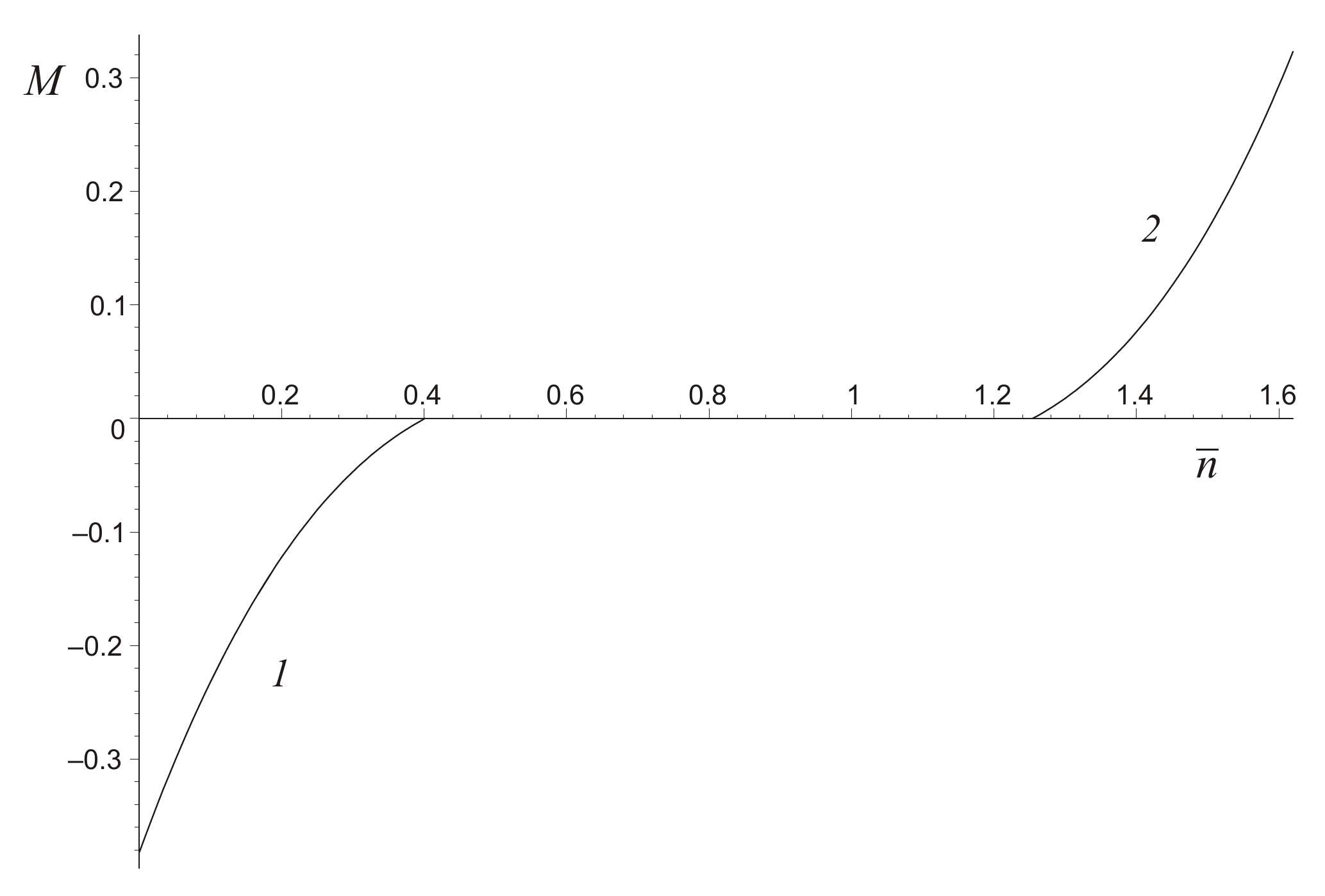}\\
  \caption{Accordance between the density of a fluid and values of the chemical potential $M$; solid line 1 -- gaseous phase; 2 -- liquid phase.}\label{fig_6}
 \end{centering}
\end{figure}

In case of temperatures below $T_{c}$ dependence of the chemical potential on density has the form represented in Figure~\ref{fig_6}. Moreover
 parts of the curve signed as 1 and 2 correspond to maximum of the expression $E(\bar{\rho}_{0})$ as it is depicted in Figure~\ref{fig_1}. The
 solution $\bar{\rho}_{01}$  matches the positive values of $M$, and $\bar{\rho}_{03}$  \-- the negative ones.
Since we use the method of steepest descent, other solutions $\bar{\rho}_{0n}$ do not contribute to the calculation of $E(\bar{\rho}_{0})$.

Thus in temperature region above the critical one $T_{c}$ the chemical potential $M$ is a smooth and continuous function of density, as it can be seen in Figure~\ref{fig_3}. At temperatures $T < T_{c}$  the curve of the chemical potential $M$ has a gap when $M$ change its sign, as it can be seen in Figure~\ref{fig_6}. Herewith negative values of $M$ correspond to the region of low densities (relatively to $n_{c}$), and positive values of $M$ describe the region of high densities. The chemical potential (in our case the effective chemical potential $M$ proportional to $\beta \mu$) is a controlling parameter in the grand canonical distribution. This value determines the average number of particles at some fixed temperature for given interaction potential between particles. Thus the systems density can be determined unambiguously by changing the values of $M$.
\begin{figure}[t!]
\begin{centering}
\includegraphics[width=200pt]{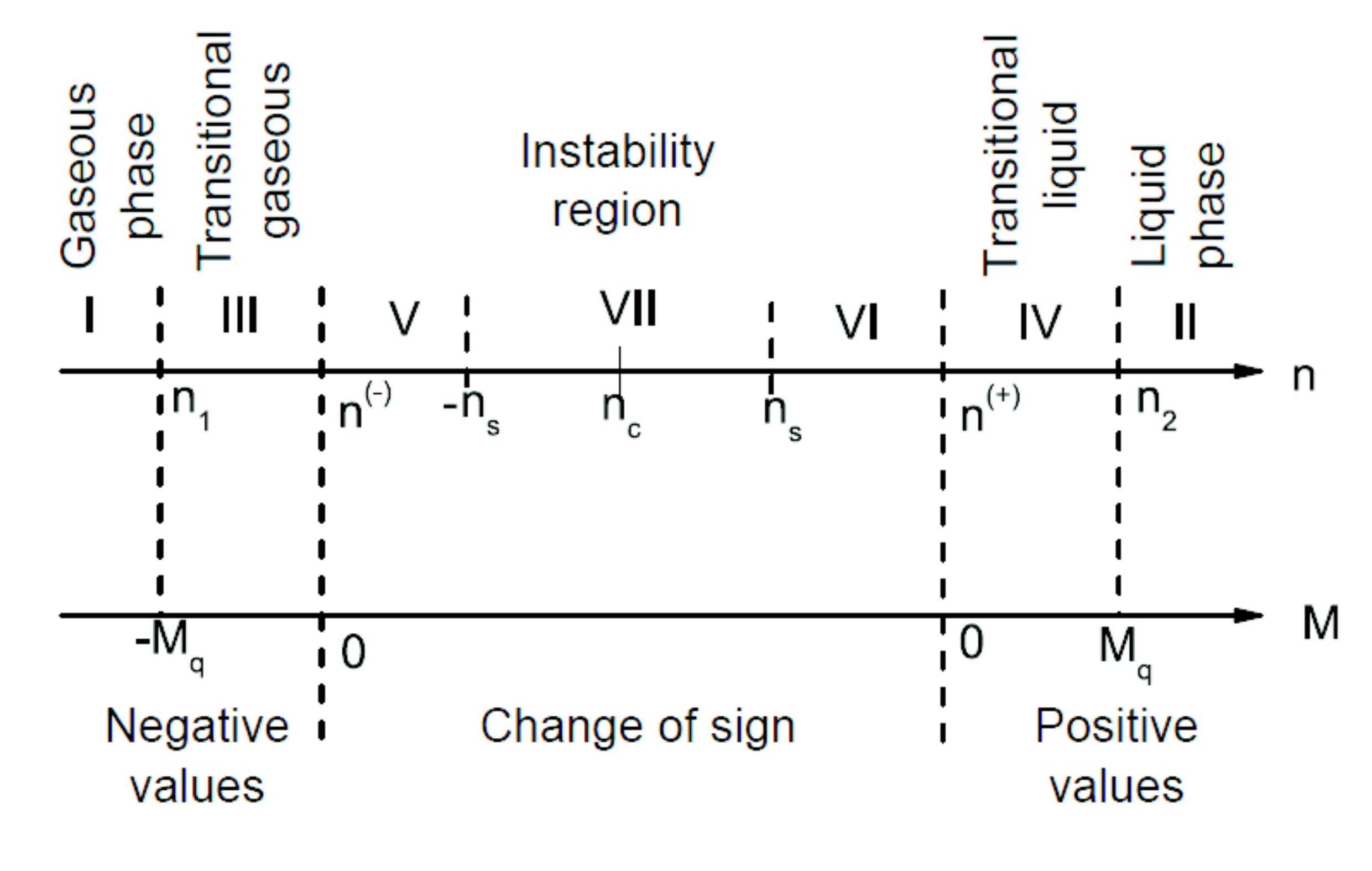}
\caption{Relation between density ranges of a simple fluid at $T<T_c$ and the chemical potential values $M$.}\label{fig_7}
\end{centering}
\end{figure}

The relation between values of $M$ and densities of the system calculated at $T < T_{c}$ according to the Eq. (\ref{eq_M_n}) is represented in Figure~\ref{fig_7}. There are several singular values of $M$. The former is $|M| = M_{q}$ (\ref{MQ}):
\begin{itemize}
              \item at $M = -M_{q}$ the density of the system is defined by the expression
              %equation (6.9)
\begin{equation}\label{n1}
    n_{1}=n_{c}-n_{g},
\end{equation}
where
%equation (6.10)
\begin{equation}\label{nG}
    n_{g}=\left( -\frac{8\tilde{a}_{2}}{a_{4}}\frac{\tau}{\tau+1}\right)^{\frac{1}{2}};
\end{equation}

              \item the case $M = M_{q}$ leads to
              %equation (6.11)
\begin{equation}\label{n2}
    n_{2}=n_{c}+n_{g}.
\end{equation}
The values of $n_{1}(T)$ and $n_{2}(T)$ are represented in Figure~\ref{fig_7}.

              \item for all $|M| > M_{q}$ the single solution $\bar{\rho}_{0}$ of the equation Eq. \eqref{eq_for_ro} exists;
              \item the region of values $M < -M_{q}$ correspond to the densities of the gaseous phase ($\bar{n} < n_{1}$);
              \item the region of values $M > M_{q}$ describe the liquid phase ($\bar{n} > n_{2}$).
            \end{itemize}
The latter characteristic value of the chemical potential is $M=0$.
In the limit of $M$ approaching to zero from below the density takes on the following value
%equation (6.12)
\begin{equation}\label{n_minus}
n^{(-)} = \lim_{M\rightarrow - 0}\bar n = n_c + \lim_{M\rightarrow - 0} \bar\rho_{03} = n_c - n_t,
\end{equation}
here
%equation (6.13)
\begin{equation}\label{nt}
n_t = \left( - \frac{6\tilde a_2\tau}{a_4(1+\tau)} \right)^{1/2}.
\end{equation}
If $M$ approaches to zero from above, then
%equation (6.14)
\begin{equation}\label{n_plus}
n^{(+)} = \lim_{M\rightarrow +0}\bar n = n_c + \lim_{M\rightarrow +0} \bar\rho_{01} = n_c +  n_t.
\end{equation}

The first order phase transition occurs when the chemical potential turns into zero. The density jump appears in the region of temperatures below the critical one $T_{c}$, where $M$ changes its sign
%equation (6.15)
\begin{equation}\label{nr_delta}
\Delta \bar n = n^{(+)} - n^{(-)} = 2 n_t.
\end{equation}
At temperatures $T > T_{c}$ such jump is absent as the equation Eq. (\ref{eq_for_ro}) has a single solution $\bar{\rho}_{0}$, which is a continuous function of $M$.

The state equation in this case takes on the form
%equation (6.16)
\begin{equation}\label{state_eq_4}
\frac{P v}{k T} = \left[ f + \frac{\tilde a_2}{2} \frac{\tau}{1+\tau}n_c^2 + \frac{a_4}{6} \bar n (\bar n - n_c)^3  - \frac{a_4}{24} (\bar n - n_c)^4  \right]
\left[  \Theta (n^{(-)} - \bar n) +  \Theta (\bar n - n^{(+)}) \right],
\end{equation}
where the quantities $n^{(-)}$ and $n^{(+)}$ are defined in Eq. (\ref{n_minus}) and Eq. (\ref{n_plus}).

Note that in temperature region $T > T_{c}$ the following equality is valid
%equation (6.17)
\begin{equation}\label{np_nm_nc}
    n^{(-)} = n^{(+)} = n_{c}
\end{equation}
and the equation Eq. (\ref{state_eq_4}) turns to be identical to the equation Eq. (\ref{state_eq_3}), that describes the region of temperatures $T > T_{c}$.

\renewcommand{\theequation}{\arabic{section}.\arabic{equation}}
\section{Conclusions}
\setcounter{equation}{0}

Using the general principles of statistical mechanics in frames of the grand canonical ensemble we propose the method to calculate the grand
partition function of a simple bulk fluid model of $N$ interactive particles. The Morse potential was chosen as an interaction potential to provide
estimations.

In course of calculating the grand partition function we used the reference system, which is formed from some part of repulsive component of
interaction potential. We found that such a choice of the reference system allows us to make summation over number of particles $N$ and integration
over their coordinates. As a result an evident form of the transition jacobian from the set of variables, which characterize individual particles,
to collective variables is obtained (some of the CV's average values are connected with the order parameter of the first order phase transition).
We introduced coefficients of the transition jacobian, which are polynomials over series of collective variables in the argument of the exponent.
The coefficients are expressed via the special functions $T_{ \! n} \! (\alpha^{ \! *}{,}p)$ , which are represented in the form of rapidly
convergent series. The arguments ($\alpha^{ \! *}$ and $p$)  of the latter special functions are real positive values: $\alpha^{ \! *}$ is related
to some fixed value of the chemical potential $\mu^{ \! *}$ and $p$ is proportional to the reference system potential.

We obtained the representation of the grand partition function of the cell model, which is general and valid both far from the critical point
and directly in its vicinity. In this model each cell can contain an arbitrary number of particles, in contrast to the Ising model.

The considered mean-field approximation is valid far away from the vicinity of the critical point. The state
equation derived in this work describes a behavior of a simple fluid system in wide temperature ranges below and above the critical
temperature $T_{ \! c}$. The curve of pressure as a function of density have horizontal parts in temperature ranges $T \!\! < \! T_{ \! c}$.
This fact particularly  describe a density jump in the first order phase transition. The curve that circumflex the rectilinear parts allows us
to obtain the binodal line.

Consideration of higher order approximations for calculating the state equation is the subject of a separate research.

\vspace{1cm}
This work was partly supported by the European Commission under the project STREVCOMS \\ {PIRSES-2013-612669}.
\vspace{1cm}

\appendix
\renewcommand{\theequation}{\arabic{section}.\arabic{equation}}
\section{Appendix A}
\setcounter{equation}{0}

Evidently, using the site representation Eq. (\ref{site}) the expression for $F(\nu)$ in Eq. (\ref{F_1}) can be factorized
%equation (A.1)
\begin{equation}\label{App1}
F(\nu) = \prod_l F_l(\nu), \tag{A.1}
\end{equation}
here
%equation (A.2)
\begin{equation}\label{App1a}
F_l(\nu) = g_{\Psi}^{\frac{1}{N_{v}}}\int^\infty_{-\infty} d \varphi_l \exp  \left[ -\frac{\varphi_{l}^{2}}{2\beta \chi\Psi(0)}\right] \sli_{m=0}^{\infty} \frac{(\alpha^*)^m}{m!} e^{-2\pi i m(\nu_l-\varphi_l/2\pi)}. \tag{A.2}
\end{equation}
Evidently, the equality $e^x=\sli_{m=0}^\infty\frac{x^m}{m!}$ is used in Eq. \eqref{App1a}. After integration over $\varphi_l$ in Eq. (\ref{F_2}) it can be expressed as follows
%equation (A.3)
\begin{equation}\label{App2}
F_l(\nu) = \sli_{m=0}^\infty \frac{(\alpha^*)^m}{m!} e^{-pm^2} e^{-2\pi i m \nu_l}, \tag{A.3}
\end{equation}
and then can be represented in the form of the cumulant expansion Eq. (\ref{F_cum})
%equation ()
\begin{equation*}
 \bar{F}_{l}(\nu)=  \exp\left[\sum \limits_{n=0}^{m_{0}} \frac{(-2\pi i)^n}{n!} \cM_n \nu_{\vl}^{n}\right],
\end{equation*}
where $m_{0} \rightarrow \infty$ and the cumulants $\cM_n$ can be found from the equalities
%equation (A.4)
\begin{equation}\label{App3}
\frac{\partial^{n} {F}_{l}(\nu)}{\partial \nu_{l}^{n}}\Bigg|_{\nu_{l}=0} =\frac{ \partial^{n} \bar {F}_l(\nu)}{\partial \nu_{l}^{n}}\Bigg|_{\nu_{l}=0}. \tag{A.4}
\end{equation}
As a result
%equation (A.5)
\begin{align}\label{cum_Mn}
&
\cM_0 = \ln T_0(\alpha^*,p),\nonumber \\
&
\cM_1 = T_1 / T_0, \quad \cM_2 = T_2 / T_0 - \cM_1^2, \nonumber \\
&
\cM_3 = T_3 / T_0 - \cM_1^3 - 3 \cM_1 \cM_2, \nonumber \\
&
\cM_4 = T_4 / T_0 - \cM_1^4 - 6 \cM_1^2 \cM_2 - 4 \cM_1 \cM_3 - 3 \cM_2^2, \nonumber \\
&
\cM_5 = T_5 / T_0 - \cM_1^5 - 10 \cM_1^2 \cM_3- 10 \cM_1^3 \cM_2 - 15 \cM_1 \cM_2^2 - 5 \cM_1 \cM_4 - 10 \cM_2 \cM_3, \nonumber \\
&
\cM_6 = T_6 / T_0 - \cM_1^6 - 15 \cM_1^4 \cM_2 - 20 \cM_1^3 \cM_3 - 15 \cM_1^2 \cM_4 - 45 \cM_1^2 \cM_2^2 - \nonumber \\
&
- 60 \cM_1 \cM_2 \cM_3 - 6 \cM_1 \cM_5  - 15 \cM_2^3 - 15 \cM_2\cM_4 - 10 \cM_3^2. \tag{A.5}
\end{align}

Here $T_n(\alpha^*, p)$ are special functions expressed in Eq. (\ref{T_spec}).

Taking Eq. (\ref{F_cum}) into account we obtain the following expression for the transition jacobian $J(\rho)$
%equation (A.6)
\begin{equation}\label{App4}
    J(\rho)=\prod \limits_{l} J_{l}(\rho_{l}), \tag{A.6}
\end{equation}
here
%equation (A.7)
\begin{equation}\label{App5}
  J_{l}(\rho_{l})= \int_{-\infty}^\infty d\nu_{l} e^{2 \pi i \nu_{l}\rho_{l}}\exp\left[\sum \limits_{n=0}^{m_{0}} \frac{(-2\pi i)^n}{n!} \cM_n \nu_{l}^n \right]. \tag{A.7}
\end{equation}

  The Eq. \eqref{App5} contains a polynomial over degrees of the real variable $\nu_l$ in the index of the exponent. The convergency of the integral over this variable is provided by even powers. This can be easily seen if represent Eq. (\ref{App5}) in the following form
  %equation (A.8)
\begin{equation}\label{App6}
   J_{l}(\rho_{l})= 2\pi \int^\infty_{-\infty} d x e^{ i x\rho_{l}} e^{f(x)} ( \cos f_1(x) - i \sin f_1 (x)), \tag{A.8}
\end{equation}
here $x=2 \pi \nu_{l}$ and
%equation (A.9)
\begin{align}\label{App7}
& f(x) = - \frac{1}{2} \cM_2 x^2 + \frac{1}{4!} \cM_4 x^4 - \frac{1}{6!} \cM_6 x^6,\nonumber \\
& f_1(x) = \cM_1 x - \frac{1}{3!} \cM_3 x^3 + \frac{1}{5!} \cM_5 x^5. \tag{A.9}
\end{align}
Here $m_0=6$ is assigned to make the Eq. \eqref{App6} definite, so to say the approximation used in~\cite{jukhn_tmf} is applied.

The result of integration over variables $\nu_l$ in Eq. (\ref{App6}) can be represented in the following form
%equation (A.10)
\begin{equation}\label{App8}
    \bar{J}_{l}(\rho_{l})=\exp\left[-\sum \limits_{n=0}^{n_{0}} \frac{a_{n}}{n!}  \rho_{l} ^{n} \right], \qquad n_{0} \rightarrow \infty. \tag{A.10}
\end{equation}

\begin{figure}[t!]
 \begin{centering}
  \includegraphics[width=200pt]{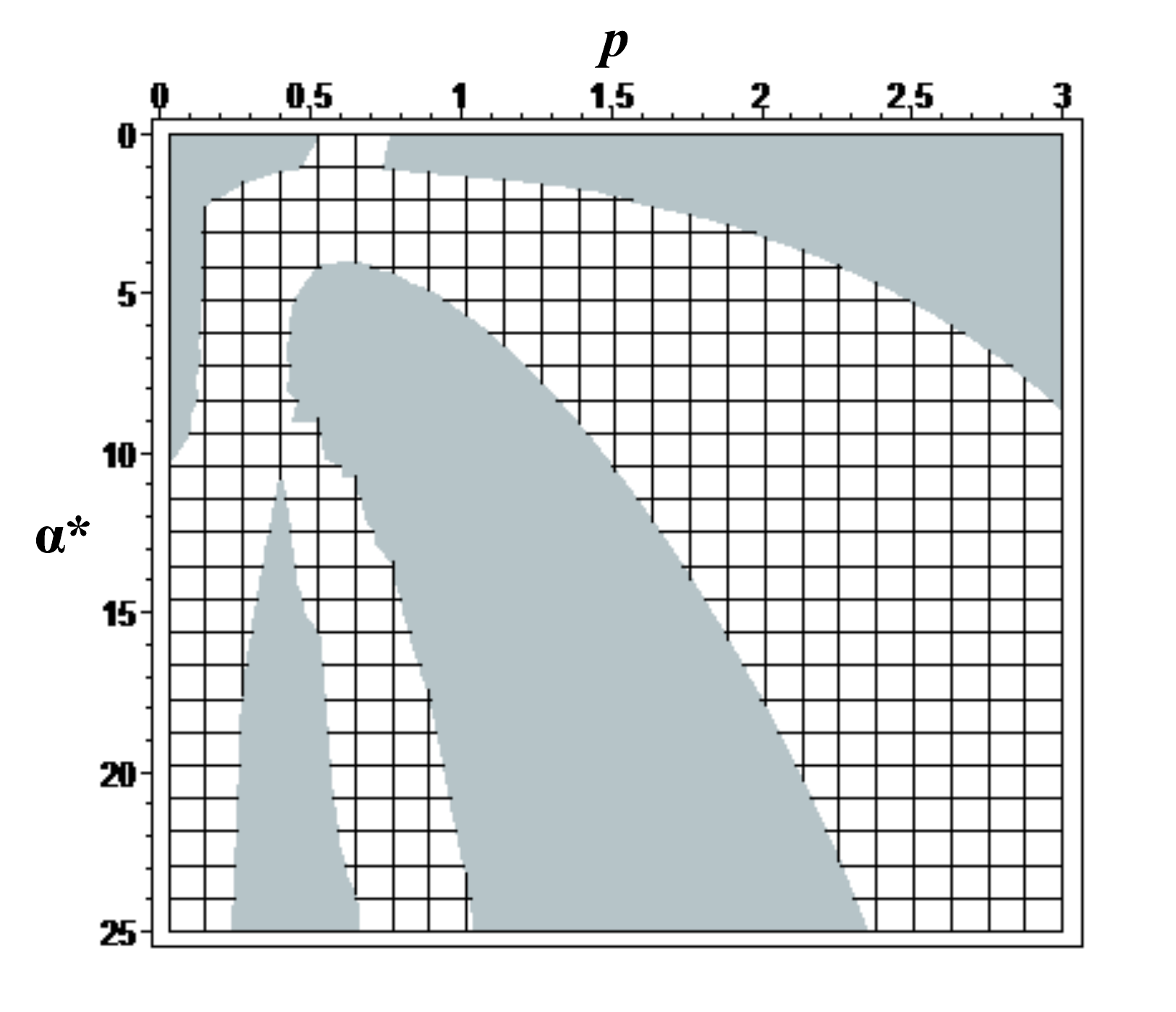}\\
  \caption{Regions of the cumulant values $\cM_{6}>0$ (signed by white color) depending on the values $\alpha^{*}$ and $p$.}\label{fig_8}
  \end{centering}
\end{figure}

 The infinite series in Eq. (\ref{App8}) will be restricted to $n_{0}=4$. Herewith, the coefficients $a_n$ are real values and can be expressed by the relations
%equation (A.11)
\begin{align}\label{App9}
& a_0 = \ln(2\pi) - \ln I_0, \quad a_1 = -J_1/I_0, \quad a_2 = I_2/I_0 + a_1^2,\nonumber \\
& a_3 = J_3 / I_0 - a_1^3 + 3 a_1 a_2, \tag{A.11} \\
& a_4 = -I_4/I_0 + a_1^4 - 6 a_1^2 a_2 + 4 a_1 a_3 + 3 a_2^2, \nonumber
\end{align}
here such notions are used
%equation (A.12)
\begin{align}\label{App10}
& I_n = \int_{-\infty}^\infty  d x x^n \cos(f_1(x)) e^{f(x)},\nonumber \\
& J_n = \int_{-\infty}^\infty  d x x^n \sin(f_1(x)) e^{f(x)}. \tag{A.12}
\end{align}
As it was mentioned above, convergency of the integrals that are expressed in Eq. (\ref{App10}) occurs for all values of
%equation (A.13)
\begin{equation}\label{condition M}
    \cM_{2}>0, \quad \cM_{4}<0, \quad \cM_{6}>0. \tag{A.13}
\end{equation}
The condition $\cM_{6}>0$ is sufficient for existence of quantities $I_{n}(\alpha^{*},p)$ and
$J_{n}(\alpha^{*},p)$. The results of the calculations show that the value of $\cM_{2}>0$ for any values
of $\alpha^{*}$ and $p$. Note that $z^{*}=e^{\beta_{c}\mu^{*}}$ and $p$ from Eq. (\ref{p})
take on real positive values.

The cumulants $\cM_{4}$ and $\cM_{6}$ are real, but they may take on both positive and negative values.
 Easy to see, that there is a region of values $0<\alpha^{*}<25$ and $0{,}1<p<3$, which satisfy the condition Eq. (\ref{condition M}). Here $\cM_{4}<0$ both with $\cM_{6}>0$ allow us to find corresponding values of $a_{n}$ (see Figure~\ref{fig_8}).

An example of the cumulants values $\cM_{n}$ and accordant coefficients $a_{n}$ in
case of $R_{0}/\alpha= 3 ln~2$, $\alpha^{*}=15$, $p=0{,}8$ and $\chi=1{,}695$ from Eq. (\ref{p}) (see Sect. 3) is given below.
%equation (A.14)
\begin{align}\label{an_Mn_value}
&
    \cM_{0}=3,6996, \quad \cM_{1}=1,3484, \quad  \cM_{2}=0,4525, \quad  \cM_{3}=0,0418, \quad    \cM_{4}=-0,0146, \nonumber \\
    &
    \cM_{5}=-0,0339, \quad   \cM_{6}=0.0212, \tag{A.14} \\
    &
    a_{0}=-1,1050, \quad   a_{1}=-3,6552, \quad    a_{2}=4,4291, \quad    a_{3}=-4,5509, \quad
    a_{4}=5,4961.  \nonumber
\end{align}

\renewcommand{\theequation}{\arabic{section}.\arabic{equation}}
\section{Appendix B}
\setcounter{equation}{0}

\begin{figure}[t!]
 \begin{centering}
    \includegraphics[width=200pt]{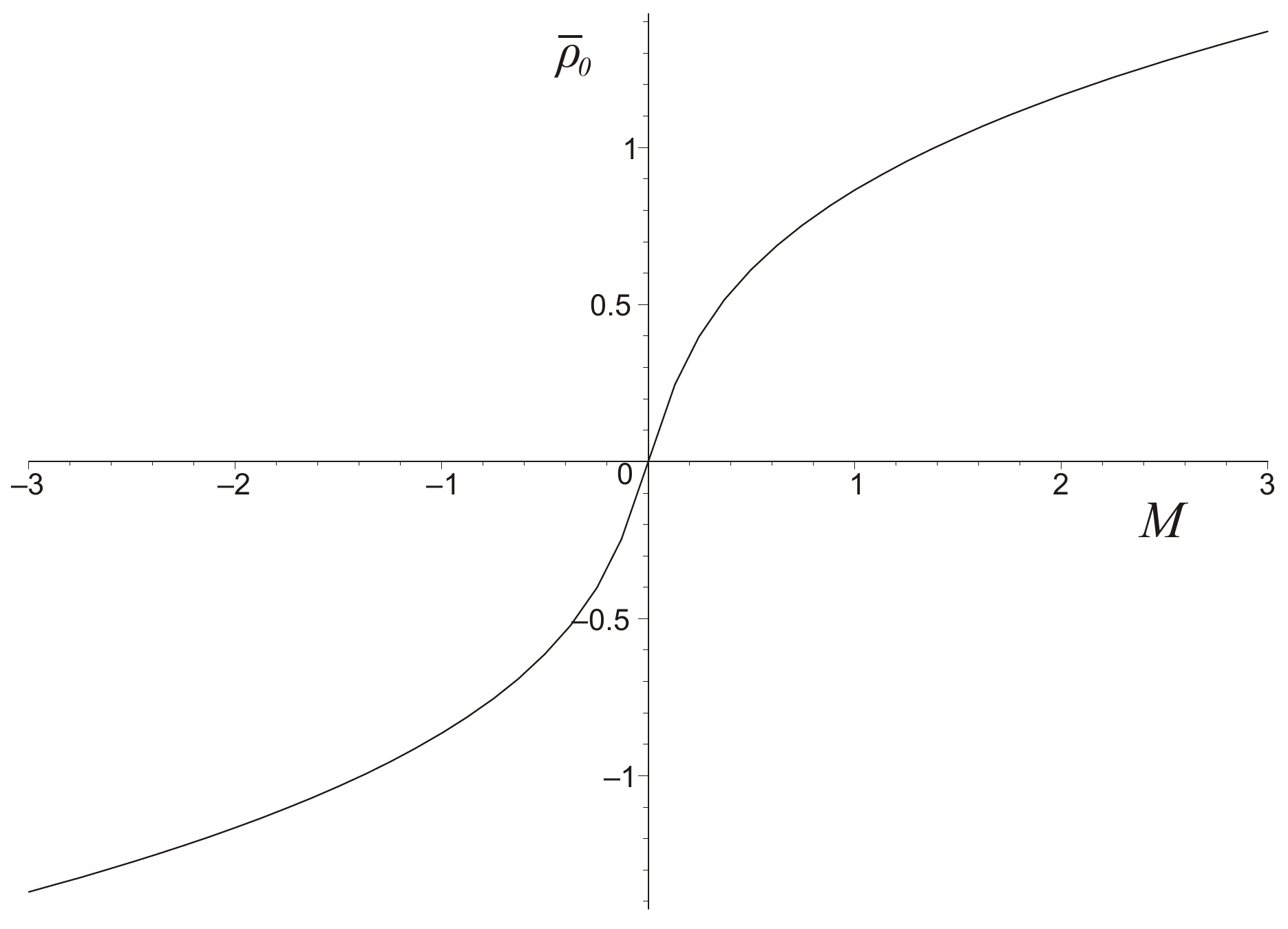}\\
  \caption{Dependence of the solution $\bar\rho_0$ on the chemical potential $M$ at $T>T_c$ in case of $R_0/\alpha=3 ln~2$, $\alpha^{\star}=15$ and $p=0{,}8$.}\label{fig_9}
\end{centering}
\end{figure}

According to Eq. (\ref{GPF_7}) the expression of the GPF at $T>T_{c}$ in the mean-field approximation has
the form
%equation (B.1)
\begin{equation}\label{GPF_8c}
  \ln\Xi_0 = N_{v} \left( \cM_0 - a_0 + E_0(\mu) + E(\bar\rho_0) \right), \tag{B.1}
\end{equation}
here
%equation (B.2,B.3)
\begin{align}\label{def_E0r_E0M}
    &E(\bar{\rho}_0)= M \bar\rho_0 - \frac{1}{2}\tilde{d}(0) \bar\rho_0^2 - \frac{a_{4}}{24}\bar\rho_0^4, \tag{B.2} \\
    &E_{0}(\mu)= M n_c + \frac{1}{2}\tilde{d}(0) n_c^2 + \frac{a_{4}}{24} n_c^4, \tag{B.3}
\end{align}
and $\bar{\rho}_{0}$ is the solution of the equation Eq. (\ref{eq_for_ro}), which is written in reduced form as
%equation (B.4)
\begin{equation}\label{eq_ro_red}
    \bar{\rho}_{0}^{3} + p\bar{\rho}_{0} + q = 0, \tag{B.4}
\end{equation}
here
%equation (B.5)
\begin{equation}\label{def_p_q}
    p = \frac{6\tilde{d}(0)}{a_{4}}, \quad q = - \frac{6M}{a_{4}}. \tag{B.5}
\end{equation}
Solution unicity of this equation is provided by the positive discriminant (here $\tilde{d}(0)>0$)
%equation (B.6)
\begin{equation}
    Q = \left( \frac{2\tilde{d}(0)}{a_{4}}\right)^{3} + \left( \frac{3M}{a_{4}}\right)^{2}. \tag{B.6}
\end{equation}
Note, that the sign of $Q$ is not dependent on the sign of the chemical potential. Among three existing solutions there is one that is real
%equation (B.7)
\begin{equation}\label{B_7}
    \bar{\rho}_{0} = \left( \frac{3M}{a_{4}} + \sqrt{Q} \right)^{1/3} +\left( \frac{3M}{a_{4}} - \sqrt{Q} \right)^{1/3}. \tag{B.7}
\end{equation}
The Eq. \eqref{B_7} defines a dependence of the chemical potential $M$ on density and temperature and the  diagram of $\bar{\rho}_{0}$ as a function of $M$ at $T>T_c$ is presented in Figure~\ref{fig_9}.

\begin{figure}[t!]
\centerline{\includegraphics[width=200pt]{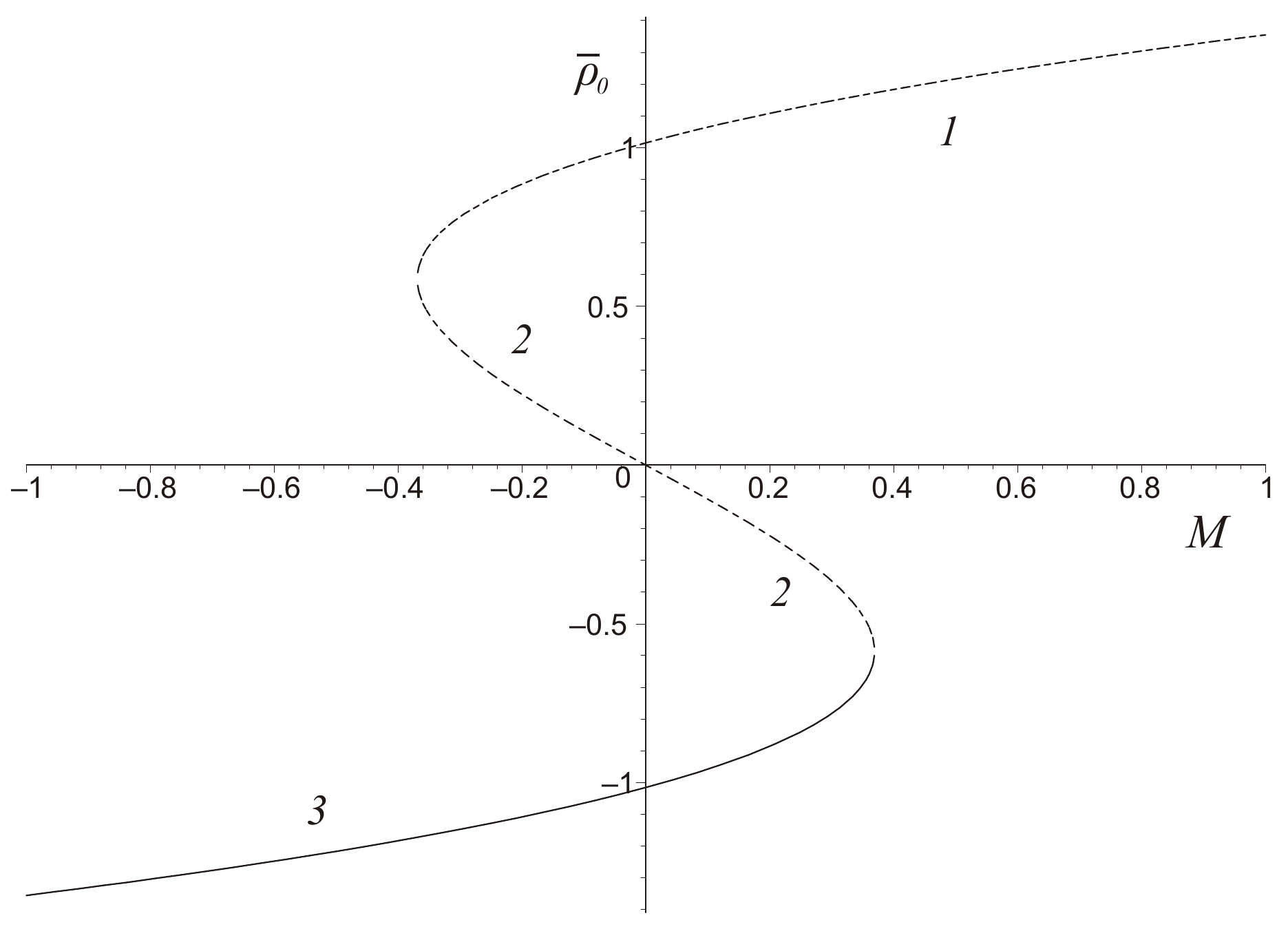}}
\caption{Dependence of the roots (\ref{eq_ro_red}) on the chemical potential values $|M|<M_q$ in temperature region $T > T_{c}$  (curve 1 corresponds to the
solution $\bar\rho_{01}$, curve 2 \--- $\bar\rho_{02}$, curve 3 \--- $\bar\rho_{03}$}).\label{fig_10}
\end{figure}

Calculation of the GPF Eq. (\ref{GPF_6}) in the mean-field approximation at $T<T_{c}$ leads us to the following expression
%equation (B.8)
\begin{equation}\label{GPF_9}
  \Xi_{0}=  N_{v} \left( \cM_0 - a_0 + E_0(\mu)\right) + N_{v} E(\bar\rho_{0i}), \tag{B.8}
\end{equation}
here
%equation (B.9)
\begin{equation}\label{E0r_2}
E_{0}(\bar{\rho}_{0i})= M\bar{\rho}_{0i} -\frac{\tilde{a}_2}{2} \frac{\tau}{\tau+1} \bar{\rho}_{0i}^{2} - \frac{a_{4}}{24}\bar{\rho}_{0i}^{4}, \tag{B.9}
\end{equation}
herewith $\bar{\rho}_{0i}$ are solutions of the equation
%equation (B.10)
\begin{equation}\label{eq_ro0_i}
    \bar{\rho}_{0i}^{3} + p\bar{\rho}_{0i} + q = 0, \tag{B.10}
\end{equation}
where the coefficients $p$ and $q$ are defined in Eq. (\ref{def_p_q}).
Unlike case of $T>T_{c}$ the equation Eq. (\ref{eq_ro0_i}) has more than one real root (see Figure~\ref{fig_10}). Thus the solution $\bar{\rho}_{0i}$ that correspond to the maximum of $E(\bar{\rho}_{0i})$ from Eq. (\ref{E0r_2}) should be chosen, according to conditions of the method of steepest descent which we use to calculate Eq. (\ref{GPF_9}).

As we can see from Eq. (\ref{discrim}) the marginal value of the chemical potential $|M_{q}|$, at which the equality $Q=0$ is fulfilled has the form
%equation (B.11)
\begin{equation}
    M_{q}=\frac{a_{4}}{3}\left(-\frac{2\tilde{d}(0)}{a_{4}}\right)^{\frac{3}{2}}, \tag{B.11}
\end{equation}
for all values of $|M|>M_{q}$ the discriminant $Q>0$, and so, the equation Eq. (\ref{eq_ro0_i}) has a single real solution Eq. (\ref{1Re_root}). In case of
$|M|<M_{q}$ ($Q<0$) there are three real solutions:
%equation (B.12)
\begin{align}\label{ro123}
&\bar \rho_{01} = 2   \rho_{0r} \cos{\frac{\alpha}{3}}, \nonumber \\
&\bar \rho_{02} = - 2  \rho_{0r} \cos \left( \frac{\alpha}{3} + \frac{\pi}{3}\right), \nonumber \\
&\bar \rho_{03} = - 2  \rho_{0r} \cos \left( \frac{\alpha}{3} - \frac{\pi}{3}\right), \tag{B.12}
\end{align}
here
%equation (B.13)
\begin{equation}\label{roR}
   \rho_{0r} =\left( - \frac{2\tilde{d}(0)}{a_{4}}\right)^{\frac{1}{2}}, \tag{B.13}
\end{equation}
the angle $\alpha$ is defined from the condition $\cos{\alpha}=\dfrac{M}{M_{q}}$  and equal to
%equation (B.14)
\begin{equation}\label{alpha_angle}
    \alpha=\arccos {\frac{M}{M_{q}}}. \tag{B.14}
\end{equation}

 \end{document}